\documentclass[runningheads,a4paper]{llncs}

\usepackage{amsmath}
\usepackage{fixltx2e}
\usepackage{verbatim}

\usepackage{color}

\usepackage{subfigure}
\usepackage[T1, OT1]{fontenc}
\DeclareTextSymbolDefault{\dh}{T1}

\usepackage{listings}

\usepackage{amssymb}
\usepackage{graphicx}
\usepackage[font={normalsize},labelfont=bf]{caption}
\usepackage{varwidth}
\usepackage[hyphens]{url}
\usepackage{tcolorbox}
\makeatletter
\g@addto@macro{\UrlBreaks}{\UrlOrds}
\makeatother
\PassOptionsToPackage{hyphens}{url}
\usepackage{fancyvrb}
\fvset{frame=single,samepage=true,fontfamily=courier,numbers=left,numbersep=2pt,baselinestretch=0.50,fontsize=\small}

\urldef{\mailsa}\path|{maturban, mln, mweigle}@cs.odu.edu|
\newcommand{\keywords}[1]{\par\addvspace\baselineskip
\noindent\keywordname\enspace\ignorespaces#1}

\newcommand\userinput[1]{\textbf{#1}}

\usepackage{lipsum}

\usepackage{tikz}
\usetikzlibrary{backgrounds}

\begin{document}

\mainmatter  

\title{Difficulties of Timestamping\\ Archived Web Pages}

\titlerunning{Challenges in Timestamping Archived Web Pages}

%
%
\author{Mohamed Aturban \and Michael L. Nelson \and Michele C. Weigle}
\authorrunning{Mohamed Aturban \and Michael L. Nelson \and Michele C. Weigle}

\institute{Dept of Computer Science, Old Dominion University, Norfolk, VA, 23529
\mailsa\\
\url{http://ws-dl.cs.odu.edu}}

%
%

\toctitle{Difficulties of Timestamping Archived Web Pages}
\tocauthor{Difficulties of Timestamping Archived Web Pages}
\maketitle

\setcounter{secnumdepth}{3}

\begin{abstract}
We show that state-of-the-art services for creating trusted timestamps in blockchain-based networks do not adequately allow for timestamping of web pages. They accept data by value (e.g., images and text), but not by reference (e.g., URIs of web pages). Also, we discuss difficulties in repeatedly generating the same cryptographic hash value of an archived web page. We then introduce several requirements to be fulfilled in order to produce repeatable hash values for archived web pages.

\end{abstract}

\section{Introduction}

The Internet Archive has made great efforts to capture and archive much of the web, allowing anyone to have access to prior states of web pages. We implicitly trust the archived content delivered by the Internet Archive (IA)\footnote{\url{https://archive.org}}, but with the current trend of extended use of other public and private web archives, we should consider the question of validity of archived web pages. For example, if a web page is archived in 1999 and replayed in 2017, how do we know that it has not been tampered with during those 18 years? 

\begin{figure*}
\centering 
	\subfigure[Accessing the archived page in August 2017 ($CO_2$ was $406.31$ $ppm$)]{
	\setlength{\fboxsep}{0pt}%
	\fbox{
	\includegraphics[scale=0.15]{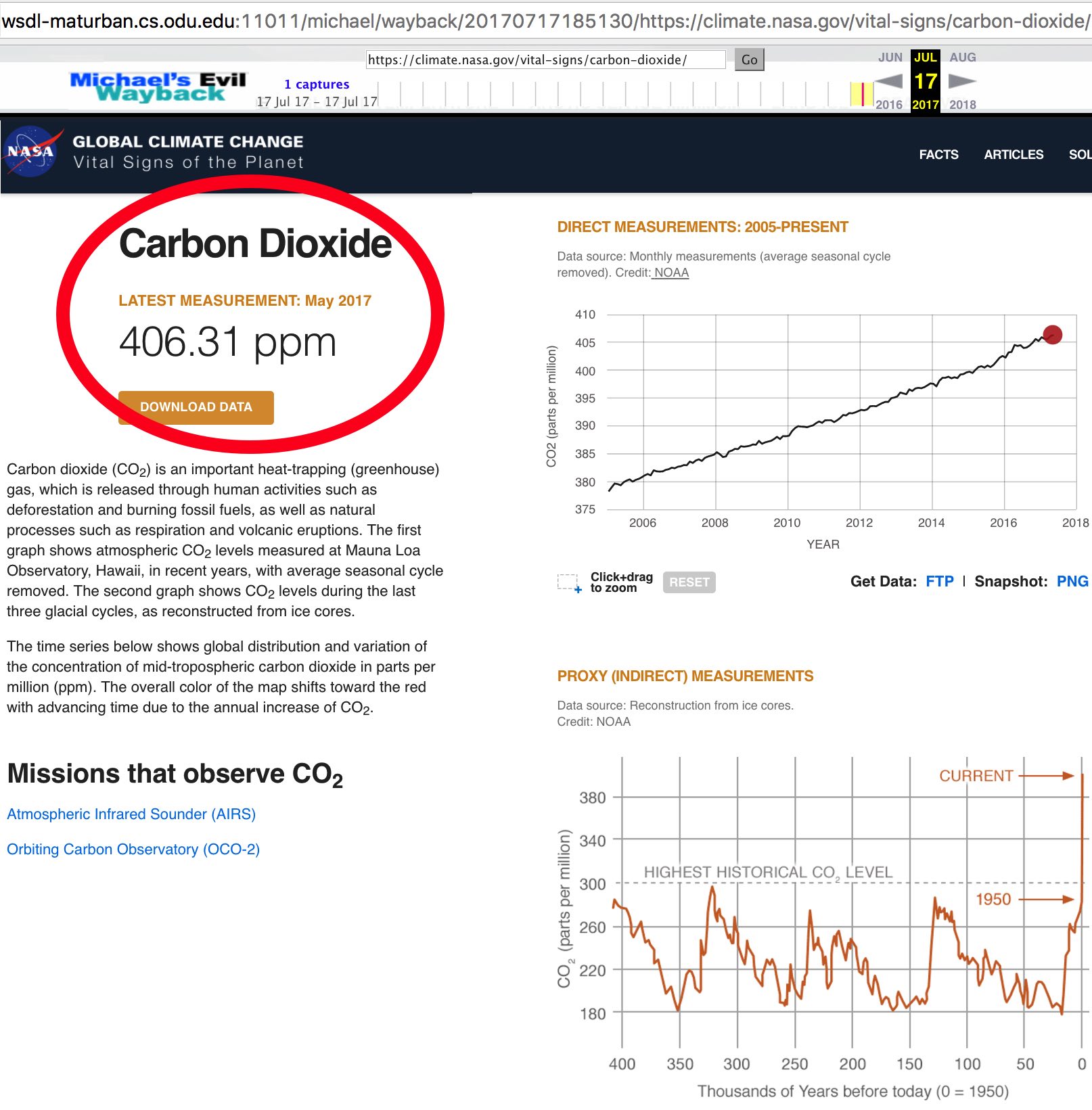}
	}
	\label{img:11011_org}
	}
	\subfigure[Accessing the same archived page in October 2017 ($CO_2$ became $270.31$ $ppm$)]{
	\setlength{\fboxsep}{0pt}%
	\fbox{
	\includegraphics[scale=0.15]{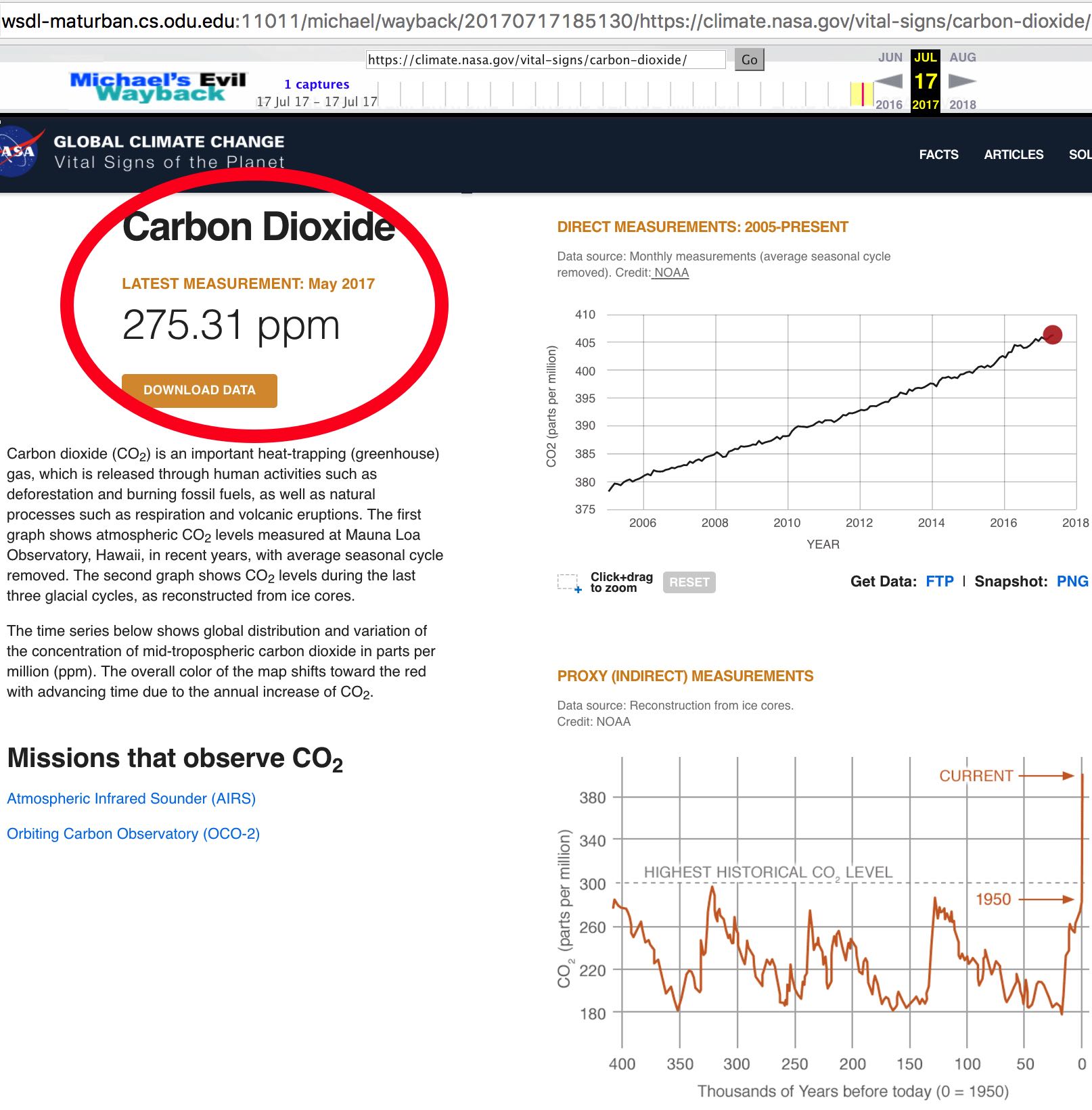}
	}
	\label{img:11011_text}
	}	
\caption{A change is made in an archived page. Which one is the real archived page? }
\label{fig:text_chnage}
\end{figure*}

When replaying the same archived web page in a web browser at different points in time, a user should be presented with the same content. Figure \ref{fig:text_chnage} shows an archived web page, or memento\footnote{A memento is an archived version of an original web page \cite{memento:rfc}.}, captured by a private web archive, ``Michael's Evil Wayback''\footnote{We established this archive to demonstrate different scenarios in this paper.}, on July 17, 2017 at 18:51 GMT. This memento is a copy of the original web page
\begin{center}
 \url{https://climate.nasa.gov/vital-signs/carbon-dioxide/}
\end{center}
Figures \ref{img:11011_org} and \ref{img:11011_text} demonstrate an unexpected result --- when replaying the memento in August 2017, the level of $CO_{2}$ (or carbon dioxide in the Earth's atmosphere) was $406.31$ $ppm$, but when revisiting the same archived page in October 2017, $CO_{2}$ became $270.31$ $ppm$. So which one is the ``real'' archived page? How can we identify whether the content, sent by the archive as a response to the most recent request, has not been tampered with? In this paper, we consider the implications of using trusted timestamping to validate archived web pages. 

Timestamping is recording the date and time of when an event occurs. For example, the HTTP Response headers ``Date'' and ``Last-Modified'' are examples of timestamps referring to different events --- ``Date'' indicates when a server generated a response message, while ``Last-Modified'' is the datetime of when the resource was last modified. A ``trusted'' timestamp is a timestamp initially created and verified by a third-party trustworthy service. Blockchain-based networks (e.g., Bitcoin \cite{nakamoto2008bitcoin,wright2015decentralized}) have been receiving increased attention recently as trustworthy systems for initiating and validating timestamps of digital documents. Once a file is timestamped in the blockchain, anyone should be able to prove the existence of the file at a particular point in time. 

In this paper, we first show that state-of-the-art timestamping services in blockchain-based networks do not allow users to submit URIs of web pages in order to establish trusted timestamps of these types of documents. Second, we discuss some difficulties in timestamping archived web pages (i.e., mementos) even if these services start accepting URIs. 

\begin{figure}[h!]
\centering
\includegraphics[width=95mm]{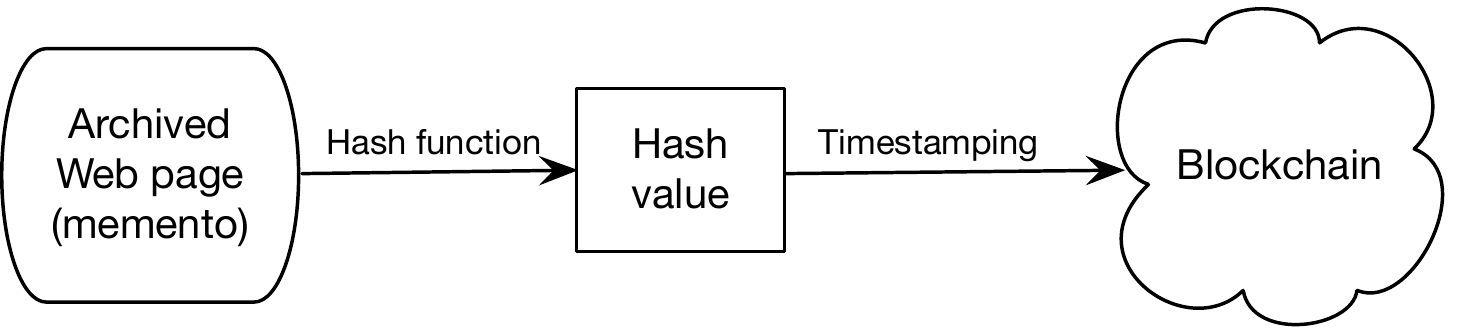}
\caption{Timestamping a hash value that summarizes a memento in the blockchain.}
\label{img:basic_timestamping}
\end{figure}

Generating a hash value on the content of a memento is one of the crucial parts in the process of timestamping the memento. As shown in Figure \ref{img:basic_timestamping}, the archived web page will not be directly timestamped in the blockchain. Instead, a hash value calculated on the content of the memento is the data to be timestamped. Thus, it is important to be able to reproduce the same hash of a particular archived web page over time. As the number of public and private web archives is increasing \cite{Kim_Nowviskie_Graham_Quon_Alliance_2017,costa2017}, we can no longer have the same level of trust in content delivered by different archives (e.g., content was tampered with in Michael's Evil Wayback as Figure \ref{fig:text_chnage} shows). It becomes essential to develop a mechanism for creating and verifying timestamps of archived web pages.

\section{Background}



\subsection{Crawling and Replaying Archived Web Pages}
\hfill\\
In order to automatically collect portions of the web, some web archives employ web crawling software, such as the Internet Archive's Heritrix \cite{sigurdhsson2010incremental,mohr2004introduction}. Having a set of seed URIs placed in a queue, Heritrix will start by fetching web pages identified by those URIs, and each time a web page is downloaded, Heritrix writes the page to a WARC file \cite{kunze2006warc}, extracts any URIs from the page, places those discovered URIs in the queue, and repeats the process. 

The crawling process will result in a set of archived pages. The Internet Archive, for example, collects over one billion web pages per week \cite{elizabethshockman2016}, and as of today, it contains 585 billion web pages \cite{iaurils}. To provide access to their archived pages, many web archives which use OpenWayback \cite{openwayback}, the open-source implementation of IA's Wayback Machine, allow users to query the archive by submitting a URI. OpenWayback will replay the content of any selected archived web page in the browser. One of the main tasks of OpenWayback is to ensure that when replaying a web page from an archive, all resources that are used to construct the page (e.g., images, style sheets, and JavaScript files) should be retrieved from the archive, not from the live web. Thus, at the time of replaying the page, OpenWayback will rewrite all links to those resources to point directly to the archive \cite{tofel2007wayback}. In addition to OpenWayback, PyWb \cite{pywb} is another tool for replaying archived web pages. It is used by Perma \cite{perma} and Webrecorder \cite{webrecorder}.

\subsection{Memento} 
\hfill\\
Memento \cite{nelson:memento:tr,memento:rfc} is an HTTP protocol extension that uses time as a new dimension to access the web by relating the current web resources to their prior states. The Memento protocol is supported by most public web archives including the Internet Archive. The protocol introduces two HTTP headers for content negotiation. First, Accept-Datetime is an HTTP Request header through which a client can request a prior state of a web resource by providing the preferred datetime (e.g., \emph{Accept-Datetime: Mon, 09 Jan 2017 11:21:57 GMT}).  Second, the Memento-Datetime HTTP Response header is sent by a server to indicate the datetime at which the resource was captured. The Memento protocol also defines the following terminology:

\begin{itemize}
\item[--] {URI-R - to identify an original resource from the live Web}
\item[--] {URI-M - to identify an archived version (memento) of the original resource at a particular point in time}
\item[--] {URI-T - a resource (TimeMap) that provides a list of mementos (URI-Ms) for a particular original resource}
\item[--] {URI-G - a resource (TimeGate) that supports content negotiation based on datetime to access prior versions of an original resource}
\end{itemize}


\subsection{The Bitcoin Blockchain} \label{bitcoin_blockchain}

Bitcoin \cite{nakamoto2008bitcoin} is a peer-to-peer electronic cash system built using the Blockchain technology \cite{wright2015decentralized}.  A ledger that contains all transactions in Bitcoin is duplicated across all nodes in the network (i.e., there is no central agency). The timestamp associated with each transaction indicates when the transaction is accepted in the Bitcoin. Services, such as OriginStamp\footnote{\url{https://originstamp.org}}, Chainpoint\footnote{\url{https://chainpoint.org/}}, and OpenTimestamps\footnote{\url{https://opentimestamps.org/}}, generate trusted timestamps in Bitcoin for digital documents. Even though timestamping steps might vary from one service to another, they follow a common procedure:
\begin{enumerate}
 \item Receiving a file, a hash, or plain text from a user
 \item Generating a hash value of received content
 \item Converting the hash to a Bitcoin address
 \item Issuing a Bitcoin transaction using the Bitcoin address as a money sender or receiver
\end{enumerate}

To verify timestamps in Bitcoin at any point in the future, the first three steps mentioned above are performed. The fourth step then would include issuing a query through the Bitcoin API to obtain information about any transactions on the given Bitcoin address. We consider the timestamp associated with the Bitcoin transaction as a trusted timestamp. Being incorruptible is the key characteristic of Bitcoin as any change in a transaction or a block requires computational power that exceeds the entire network, which is theoretically possible but unlikely to occur practically. The other important feature of Bitcoin is the decentralization of a distributed ledger which contains all transactions ever made in Bitcoin (i.e., the ledger is duplicated across all nodes). 

\section{Related Work}

Some existing work focuses on exploring security issues in web archives. Archived web pages, similar to live web pages, might change over time for different reasons, such as software or hardware upgrades, the fact that complex technologies are involved in developing web pages, and malicious attacks. For more critical archived web resources (e.g., documents acknowledged as evidence in courts), it is important to find a way to validate content served by a web archive \cite{eltgrowth2009best}.

Lerner et al. \cite{lerner2017rewriting} discovered four vulnerabilities in the Internet Archive's Wayback Machine (i.e., Archive-Escapes, Same-Origin Escapes, Archive-Escapes + Same-Origin Escapes, and Anachronism-Injection) that attackers can leverage to modify a user's view at the time when a memento is rendered in a browser. The authors suggested some defenses that could be deployed by either web archives or web publishers to prevent abusing these vulnerabilities. 

Other web archiving researchers created a shared repository in May 2017 maintained by the Harvard Library Innovation Laboratory. They use this platform to discuss potential threats in web archives. Threats would include, for instance, controlling a user's account due to Cross-Site Request Forgery (CSRF) or Cross-Site Scripting (XSS), and archived web resources reaching out to the live web. The authors provide recommendations on how to avoid such threats \cite{warcgamesgithub2017,cushman2017}. 

Eltgrowth \cite{eltgrowth2009best} outlines several judicial decisions that involve evidence (i.e., archived web pages) taken from the Internet Archive (e.g., court cases like Telewizja Polska USA, Inc. v. Echostar Satellite Corp, and St. Luke's Cataract \& Laser Institute v. James C. Sanderson). The author mentions that there is an open question whether to consider an archived web page as a duplicate of the original web page at a particular time in the past. This concern might prevent considering archived web pages as evidence. 

Yasskin \cite{webpackages2017} describes  several use cases with associated requirements for distributing copies of web packages. One use case is the authentication process, which is performed to ensure resources come from particular origins and to validate the content integrity against any attempt to tamper with or modify the content in transit. The authors did not include a use case where content might be altered at any point in time in the server.

Tools have been developed to generate trusted timestamps in blockchain-based networks. OriginStamp \cite{gipp2015decentralized} allows users to submit plain text, a hash value, or any file format (e.g., PDF/PNG files). The data is not sent to the OriginStamp's server. Instead, it is hashed in the user's browser and only the resulting hash is transmitted to the server. Once delivered, it will be added to the list of all hashes submitted by other users. Once per day, OriginStamp generates a single aggregated hash of all received hashes. This hash is then converted to a Bitcoin address that will be a part of a new Bitcoin transaction (i.e., the source or destination of a transaction in Bitcoin). The timestamp associated with the transaction is considered a trusted timestamp. OriginStamp provides an instant timestamping in the Bitcoin if a user is willing to pay a Bitcoin transaction fee. A user can verify a timestamp through OriginStamp's API or by visiting their website. The server first receives a hash from a user, then OriginStamp converts the hash to a Bitcoin address and sends a query to Bitcoin's API. If any transaction involved the given address is returned, the timestamp associated with the transaction can be used as a proof of existence. In addition to the process of verifying timestamps through OriginStamp's website, users may verify timestamps directly in Bitcoin. 

Other services, such as Chainpoint, Tangible.io\footnote{\url{http://tangible.io/en/}}, Proof of Existence\footnote{\url{https://proofofexistence.com}}, and OpenTimestamps, are based on the same concept of using Bitcoin to timestamp digital documents. Some differences between these tools include:
\begin{itemize}
\item {\emph{Cost} - The OriginStamp service can be used with no charge unless users want an instant submission to Bitcoin. Tangible.io's and Proof of Existence users, on the other hand, have to pay for the service.}
\item {\emph{Generation of aggregated hashes} - In OriginStamp, an aggregated hash is computed by storing all hashes received within a day (i.e., 24 hours)  in a file, which then will be hashed to generate a single aggregated hash. Chainpoint and OpenTimestamps uses a Merkle Tree \cite{Merkle79} to generate one aggregated hash (i.e., root hash).}
\item {\emph{The number of Bitcoin transactions v. hashes } - Services like OriginStamp, ChainPoint, and OpenTimestamps support issuing either one Bitcoin transaction per submitted hash or one transaction per aggregated hash. Other tools, such as Proof of Existence, create one Bitcoin transaction per hash.}
\item {\emph{Use} - OriginStamp, Tangible.io, and Proof of Existence provide online services through their websites that allow users to create or verify trusted timestamps using a web browser. Chainpoint and OpenTimestamps require installing client software in order to use the timestamping service.}
\item {\emph{Blockchain-based network} - Bitcoin is commonly used by all of these services to generate trusted timestamps. In addition, Chainpoint can create timestamps based on other blockchain networks like Ethereum \cite{wood2014ethereum}.}
\end{itemize}

\begin{figure}[h!]
\centering
\includegraphics[scale=0.35]{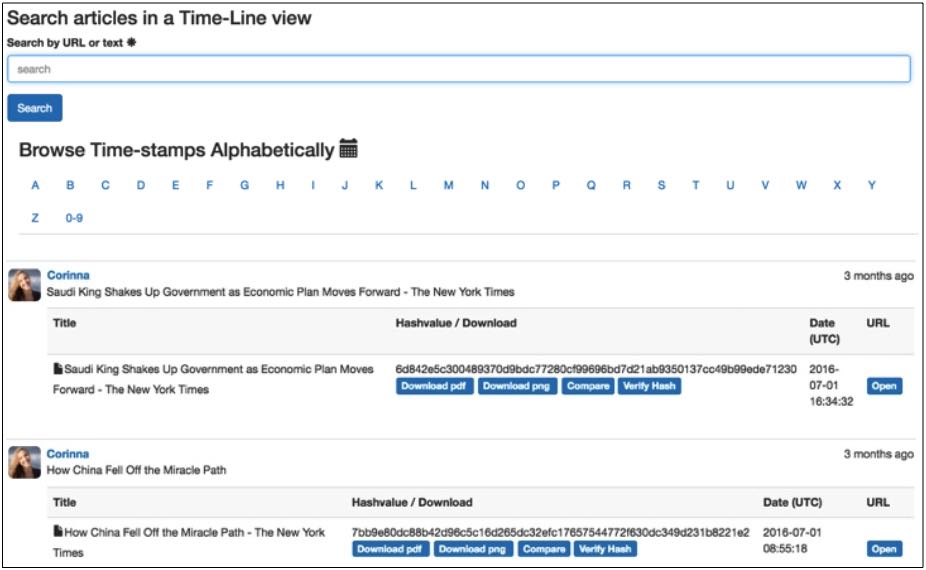}
\caption{A list of timestamped web pages. Users can search for a particular web page by typing a URI or text (from \cite{gippusingpaperforweb}).}
\label{img:originstamp_uri}
\end{figure}

Even though users of the tools mentioned above can pass data by value, such as plain text, any file format, or a hash value, they are not allowed to submit data by reference (i.e., passing a URI of a web page). In other words, these services are not directly timestamping web pages. The only exception is an additional service \cite{gippusingpaperforweb} established by OriginStamp. The service works by receiving a URI from a client and then hashing the content of the web page identified by the URI. All further steps performed to create or verify timestamps are similar to the steps mentioned in Section \ref{bitcoin_blockchain}. Figure \ref{img:originstamp_uri} (from \cite{gippusingpaperforweb}) shows the UI of this service where users can search for timestamped web pages by entering a URI. There are two disadvantages of this additional service. First, the service is no longer available on the live web\footnote{\url{https://www.isg.uni-konstanz.de/web-time-stamps/}}. Second, the hash is only generated on the HTML content of the main file identified by the URI, ignoring all embedded resources like images, scripts, and style sheets \cite{gippusingpaperforweb}. As will be illustrated in Section \ref{issues_section}, by not including embedded resources in hash calculation may leave the page vulnerable to undetected changes.

Various tools (e.g., shell scripts by Branwen \cite{gwerntimestamping2017}) calculate a hash value by considering all resources constructing a web page (e.g., images and scripts) in addition to the HTML content. This seems to be a reasonable solution for timestamping web resources, but without considering other factors, as we will see in Section \ref{issues_section}, it is difficult to produce a repeatable hash for the same web page over time.

\section{Issues in Generating Cryptographic Hashes of Mementos} \label{issues_section}

Generating a hash value on the content of a memento is the key part in the process of timestamping the memento. As shown in Figure \ref{img:basic_timestamping}, the archived web page will not be directly timestamped in blockchain-based networks (e.g., Bitcoin and Ethereum). Instead, the hash value calculated on the content is the data to be timestamped.  Regardless of the cryptographic hash function (e.g., MD5 or SHA-256), a resulting hash value should fulfill the following requirement emphasizing reproducing the same hash of a particular memento at different points in time.        
\vspace{.5cm}
\begin{center}
\tikzstyle{background rectangle}=[thin,draw=black]  
\begin{tikzpicture}[show background rectangle]
\node[align=justify, text width=10.5cm, inner sep=1em]{
If we download a memento \emph{URI-M$_x$} at time \emph{t$_n$} (denoted as \emph{URI-M$_x@$}\emph{t$_n$}), download the same memento at time \emph{t$_m$} (denoted as \emph{URI-M$_x@$}\emph{t$_m$}), and apply a hash function \emph{H} on the content of \emph{URI-M$_x@$}\emph{t$_n$} and  \emph{URI-M$_x@$}\emph{t$_m$}, then  \emph{H(URI-M$_x@$}\emph{t$_n$}\emph{)} $ = $ \emph{H(URI-M$_x@$}\emph{t$_m$}\emph{)} 
};
\node[xshift=1ex, yshift=-.7ex, overlay, fill=white, draw=white, above 
right] at (current bounding box.north west) {
\textbf{Requirement 1}: Repeatable hash values
};
\end{tikzpicture} 
\end{center}
\vspace{.4cm}
In this section, we will discuss several difficulties in generating a constant hash at different points of time for a specific archived web page. Thus, we will observe more requirements as we advance in our discussion in addition to Requirement 1 above. We will start with a simple scenario where hashes are calculated on only HTML content of mementos. The discussion then turns toward more complex scenarios encountered when all resources constructing a memento are included in the hash calculation.


\subsection{Generating hashes on HTML content only} \label{html_only}

\begin{figure}[h!]
\centering
\includegraphics[width=115mm]{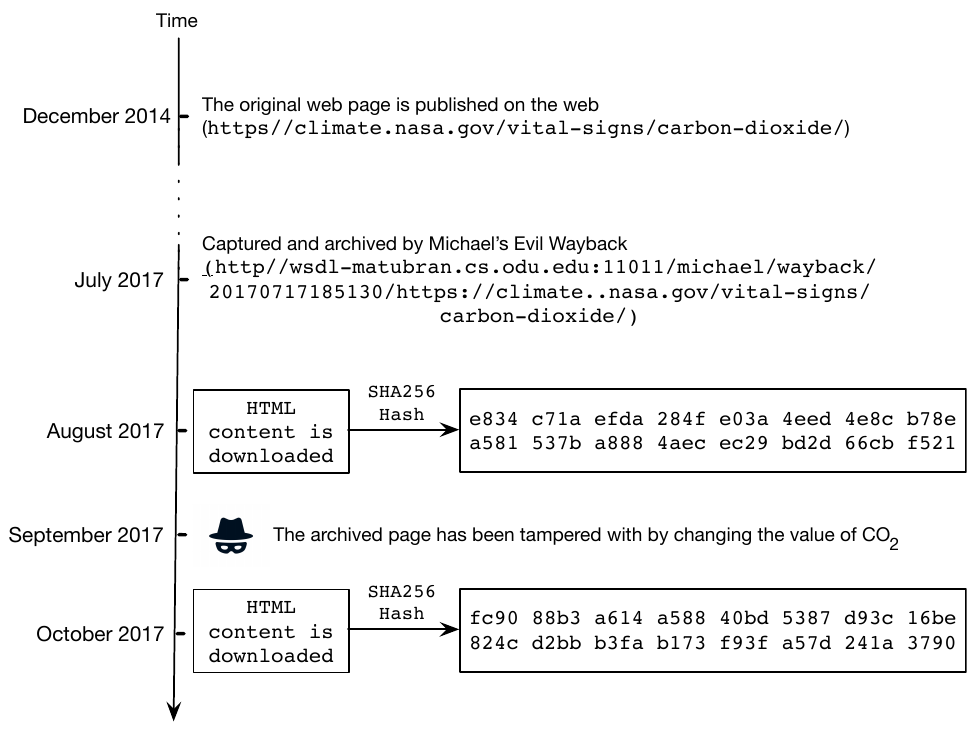}
\caption{An archived web page (http://wsdl-maturban.cs.odu.edu:11011/michael/wayback/ 20170717185130/https://climate.nasa.gov/vital-signs/carbon-dioxide/) has been tampered with, and the simple approach of generating a hash based on the HTML content successfully detected the change.}
\label{img:html_change_over_time}
\end{figure}

Consider a scenario illustrated in Figure \ref{img:html_change_over_time} where, in August 2017, a user  needs to generate a hash value based on the content of the archived page
\begin{center}
\begin{tabular}{l}
  \url{http://wsdl-maturban.cs.odu.edu:11011/michael/wayback/2017071}
  \\
  \url{7185130/https://climate.nasa.gov/vital-signs/carbon-dioxide/}
\end{tabular}
\end{center}
The memento is preserved by Michael's Evil Wayback and illustrated in Figure \ref{img:11011_org}. The user runs a ``cURL'' command as shown in Figure \ref{img:hash_on_html}, which will first download the HTML content of the main page and then generate a SHA-256 hash value of the content resulting in a hash value that ends with ``f521''. Two months later (i.e., October 2017), the user requests the same archived web page from the Michael's Evil Wayback and performs the hash calculation on the returned HTML content, observing a different hash value that ends with ``3790''. 

One possible cause of such observation of getting varied hash values is demonstrated in Figure \ref{img:html_change_over_time} with a ``black hat'' icon. Michael's Evil Wayback has tampered with the memento, in favor of individuals or organizations who deny that $CO_{2}$ is one of the main causes of global warming. The latest important $CO_2$ measurement has been changed from $406.31$ $ppm$ to $270.31$ $ppm$ as shown in Figure \ref{fig:text_chnage}. By applying the simple approach of computing hashes on the HTML content, the user becomes aware that the retrieved content in October 2017 cannot be identical to the content retrieved a couple of months earlier.   

\begin{figure}[!t]
\centering

\begin{Verbatim}[commandchars=\\\{\}]

\userinput{\%} curl -s http://wsdl-maturban.cs.odu.edu:11011/michael/wayback
/20170717185130/https://climate.nasa.gov/vital-signs/carbon-
dioxide/ | shasum -a 256

e834c71aefda284fe03a4eed4e8cb78ea581537ba8884aecec29bd2d66cbf
521 -

\end{Verbatim}
\vspace{-0.5em}
\caption{cURL command to generate a SHA-256 hash of the HTML content only.}
\label{img:hash_on_html}
\end{figure}

\begin{figure*}
\centering 
	\subfigure[Accessing the archived page in August 2017. ]{
	\setlength{\fboxsep}{0pt}%
	\fbox{
	\includegraphics[scale=0.15]{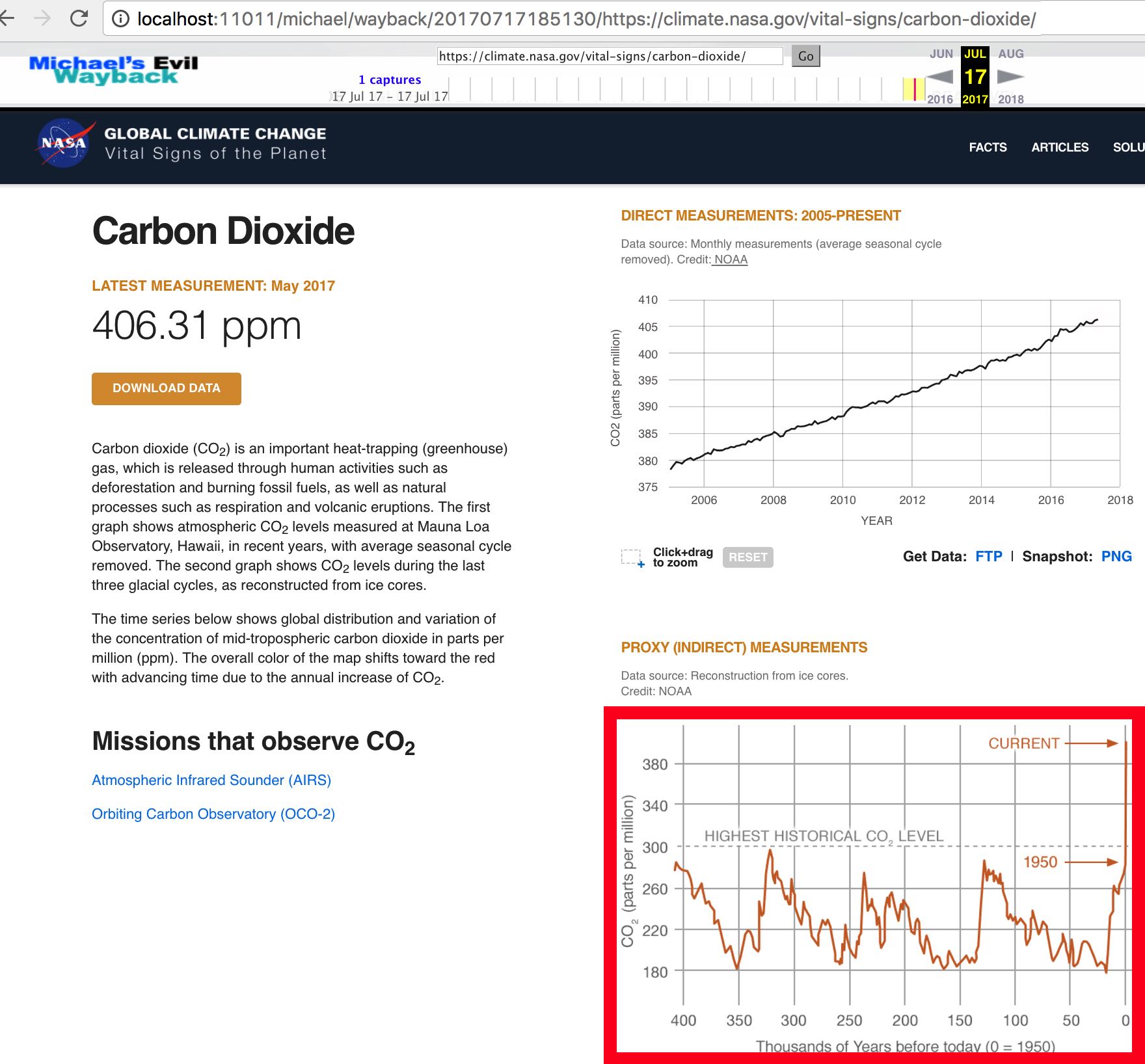}
	}
	\label{img:11011org2}
	}
	\subfigure[Accessing the archived page in October 2017]{
	\setlength{\fboxsep}{0pt}%
	\fbox{
	\includegraphics[scale=0.15]{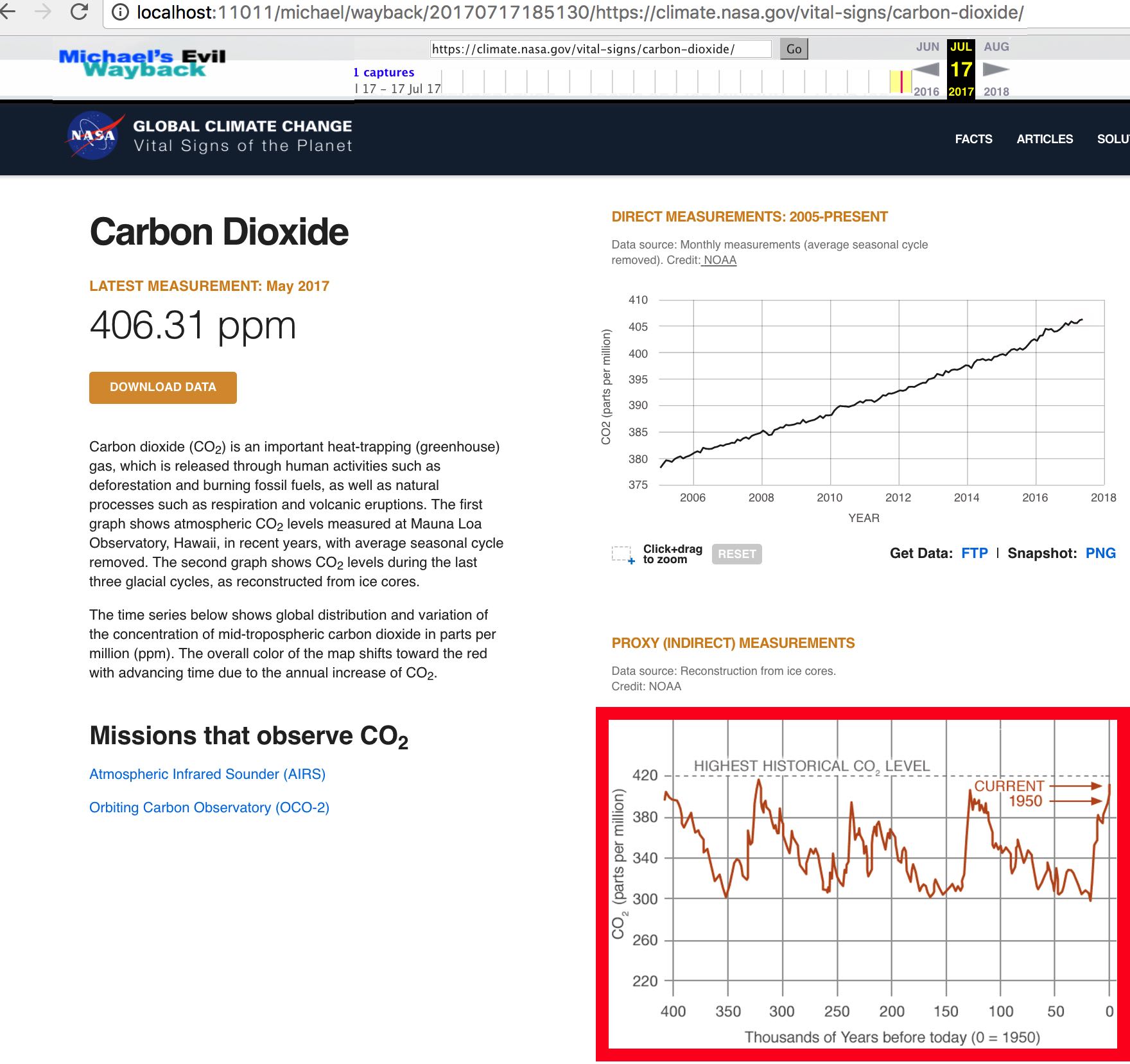}
	}
	\label{img:11011image}
	}	
\caption{A memento has been tampered with (modifying an image). The approach of hashing HTML content only does not detect the change.}
\label{fig:imagechnage2}
\end{figure*}

\begin{figure}[h!]
\centering
\includegraphics[width=115mm]{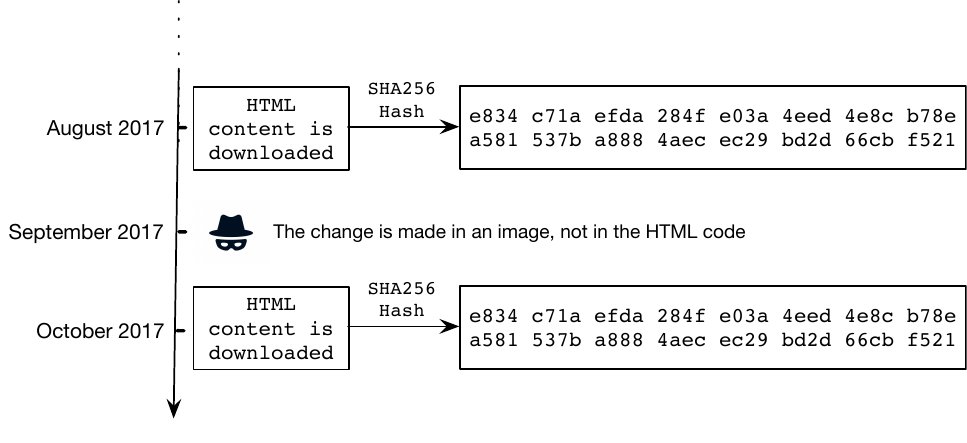}
\caption{An archived web page (http://wsdl-maturban.cs.odu.edu:11011/michael/wayback/ 20170717185130/https://climate.nasa.gov/vital-signs/carbon-dioxide/) has been tampered with, and the simple approach of generating a hash based on the HTML content alone does not detect that the archived page is corrupted.}
\label{img:image_change_over_time}
\end{figure}

We focus now on a more complicated scenario where an image or any other embedded resource constructing the archived page is altered. For instance, the bottom-right graph of the archived web page shown in Figure \ref{img:11011image} has been changing the historical records of $CO_2$. This image is located on the web at 
\begin{center}
\begin{tabular}{l}
  
\url{http://wsdl-maturban.cs.odu.edu:11011/michael/wayback/20170717}
\\
\url{185130im_/https://climate.nasa.gov/system/charts/15_co2_left_0}\\\url{61316.gif} 
  
\end{tabular}
\end{center}
It is linked from the main file of the archived page 
\begin{center}
\begin{tabular}{l}

\url{http://wsdl-maturban.cs.odu.edu:11011/michael/wayback/20170717}\\\url{185130/https://climate.nasa.gov/vital-signs/carbon-dioxide/}
  
\end{tabular}
\end{center}
Can such change be detected by the ``cURL'' command shown in Figure \ref{img:hash_on_html}? The answer is ``no'' since it only considers hashing the HTML code of the main file and not the embedded resources. Figure \ref{img:image_change_over_time} shows the results of running the command on the archived page before (Figure \ref{img:11011org2}) and after it is modified  (Figure \ref{img:11011image}). The hash values are identical, which falsely indicates that the archived page is not corrupted. Therefore, we should include embedded resources in hash calculation.

\subsection{Generating a hash of a composite memento} \label{composite}
\begin{figure}[h!]
\centering
\includegraphics[width=115mm]{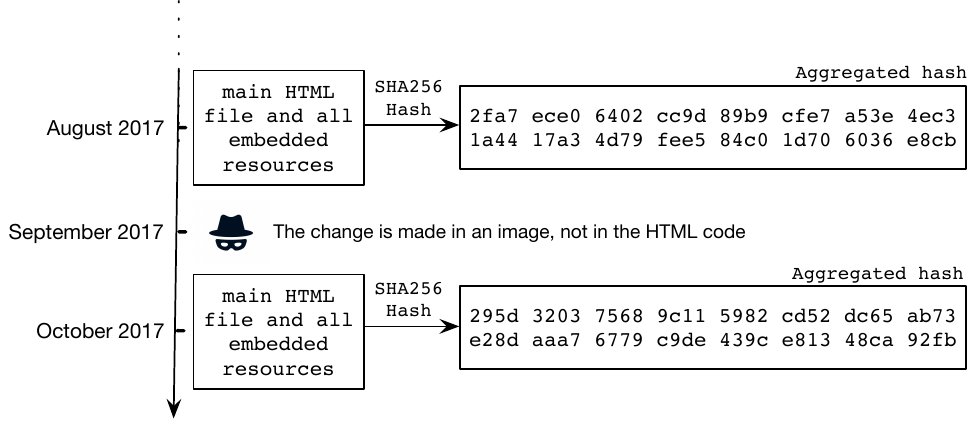}
\caption{An archived web page (http://wsdl-maturban.cs.odu.edu:11011/michael/wayback/ 20170717185130/https://climate.nasa.gov/vital-signs/carbon-dioxide/) has been tampered with. The shell script sha256\_include\_all.sh successfully detects that the archived page is corrupted.}
\label{img:image_change_over_time2}
\end{figure}

A composite memento refers to all embedded resources that comprise a memento \cite{ainsworth2014framework}. We modified the shell script (see Figure \ref{img:hash_embedded}) written by Gwern Branwen \cite{gwerntimestamping2017}. The modified script, \emph{sha256\_include\_all.sh}, computes a final hash by hashing a text file containing a set of hash values of all embedded resources constructing a memento (i.e., a composite memento). Figure \ref{img:hash_embedded_on_orig} shows an example of running this script on the content of a real archived page. Figure \ref{img:image_change_over_time2} shows the results of computing the hash on the original archived page (Figure \ref{img:11011org2}) and after an image within the memento is modified (Figure \ref{img:11011image}). The new script successfully produces two different hash values. The first one ends with ``6e8cb'' while the second hash ends with ``a92fb''. This indicates that the memento has been tampered with or altered. Thus, fulfilling Requirement 2 is essential when computing hashes.  

\begin{figure}[!t]
\centering
\begin{Verbatim}[commandchars=\\\{\}]

#!/bin/bash
#set -euo pipefail

rm -rf ~/tmp_www/*
cd ~/tmp_www/

USER_AGENT="Firefox 6.4"

FILE=\$(nice -n 20 wget --continue --unlink --page-requisites
             --timestamping -e robots=off -k  
             --user-agent="\$USER_AGENT" "\$1" 2>&1 \
             
             | egrep 'Saving to: ‘.*’' 
             | sed -e 's/Saving to: ‘//' | tr -d '’')

let "c=0"

for TARGET in \$FILE; do
 if [ -f "\$TARGET" ]; then
  let "c++"
  CONT=\$(cat \$TARGET)
  HASH=\$(echo "\$CONT" | shasum -a 256 | awk '{print \$1;}')
  echo "\$HASH" >> "allhashes.txt"
 fi
done

if [ \$c = 1 ]; then
   FINAL_HASH="\$HASH"
else
   FINAL_HASH=\$(cat "allhashes.txt" | shasum -a 256
                                    | awk '{print \$1;}')
fi

echo "Final hash: \$FINAL_HASH"


\end{Verbatim}
\vspace{-0.5em}
\caption{A shell script (sha256\_include\_all.sh) to generate a final hash by aggregating all hash values of the embedded resources in a single temporary file and hashing the file.}
\label{img:hash_embedded}
\end{figure}

\begin{figure}[!t]
\centering
\begin{Verbatim}[commandchars=\\\{\}]

\userinput{\%} sha256_include_all.sh http://wsdl-maturban.cs.odu.edu:11011/
michael/wayback/20170717185130/https://climate.nasa.gov/vital-
signs/carbon-dioxide/

Final hash: 2fa7ece06402cc9d89b9cfe7a53e4ec31a4417a34d79fee584c
01d706036e8cb

\end{Verbatim}
\vspace{-0.5em}
\caption{An example of generating an aggregated hash using sha256\_include\_all.sh.}
\label{img:hash_embedded_on_orig}
\end{figure}

\vspace{.5cm}
\begin{center}
\tikzstyle{background rectangle}=[thin,draw=black]  
\begin{tikzpicture}[show background rectangle]
\node[align=justify, text width=10.5cm, inner sep=1em]{
We should hash a composite memento. In most cases this would include hashing the main HTML file as well as other embedded resources in the memento, such as images, style sheets, JavaScript files, iframes, and others.   
};
\node[xshift=1ex, yshift=-.7ex, overlay, fill=white, draw=white, above 
right] at (current bounding box.north west) {
\textbf{Requirement 2}: Hash a composite memento
};
\end{tikzpicture} 
\end{center}

Although including embedded resources of a memento in a hash calculation may help identify memento tampering (Requirement 2), it raises more questions about whether to exclude some of those resources (e.g., archive-specific resources) for several reasons explained in the next section.

\subsection{Excluding archive-specific content} \label{archive_specific}
Before sending any requested memento to a client, archives not only insert extra code for usability (e.g., the IA's banner) in the original content of mementos but may also apply some transformation to appropriately replay content in a user's browser. An archive performs such transformations for different reasons. First, all links to embedded resources constructing an archived page are rewritten so that these resources are retrieved from the archive, not from the live web. For instance, the memento
\begin{center}
\begin{tabular}{l}
  \url{https://web.archive.org/web/20170705002324/http://www.weeklysta} 
  \\
  \url{ndard.com/}
\end{tabular}
\end{center}

{\raggedleft{}contains the logo image}

\begin{center}

  \url{http://www.weeklystandard.com/media/images/logo.png}

\end{center}

{\raggedleft{}This link to the logo image is rewriten by the Wayback Machine to point to the archive}

\begin{center}
\begin{tabular}{l}
  \url{https://web.archive.org/web/20170705161539im_/http://www.weekly}
  \\
  \url{standard.com/media/images/logo.png}
\end{tabular}
\end{center}

Another purpose of such archive-specific content is to inform users that what they are viewing is actually from an archive rather than the live web. The Internet Archive, for example, adds HTML comments at the end of the main HTML file of a memento to indicate when the memento was created and retrieved (Figure \ref{fig:ia_added_code}). In addition, archives insert content to convey information such as the archive name, current datetime, and copyright-related statements. Jones et al. \cite{jones2016rules,jonesraw20162,jonesraw2016} explore transformation of original content performed by different archives and introduce several rules for acquiring mementos and extracting text from the content.

\begin{figure}[!t]
\centering
\begin{Verbatim}[commandchars=\\\{\}]
<html>
<head> ... </head>
<body>
...
<!--
 FILE ARCHIVED ON 23:42:15 Apr 6, 2017 AND RETRIEVED FROM THE
 INTERNET ARCHIVE ON 3:40:16 Apr 7, 2017.
 JAVASCRIPT APPENDED BY WAYBACK MACHINE, COPYRIGHT INTERNET 
 ARCHIVE.

 ALL OTHER CONTENT MAY ALSO BE PROTECTED BY COPYRIGHT (17 U.S.
 C.SECTION 108(a)(3)).
-->	
</body>
</html>
\end{Verbatim}
\vspace{-0.5em}
\caption{HTML comments added by the Internet Archive.}
\label{fig:ia_added_code}
\end{figure}

\begin{figure*}
\centering 
	\subfigure[A Memento from Internet Archive
	]{
	\setlength{\fboxsep}{0pt}%
	\fbox{
	\includegraphics[width=2.7in, height=2.5in]{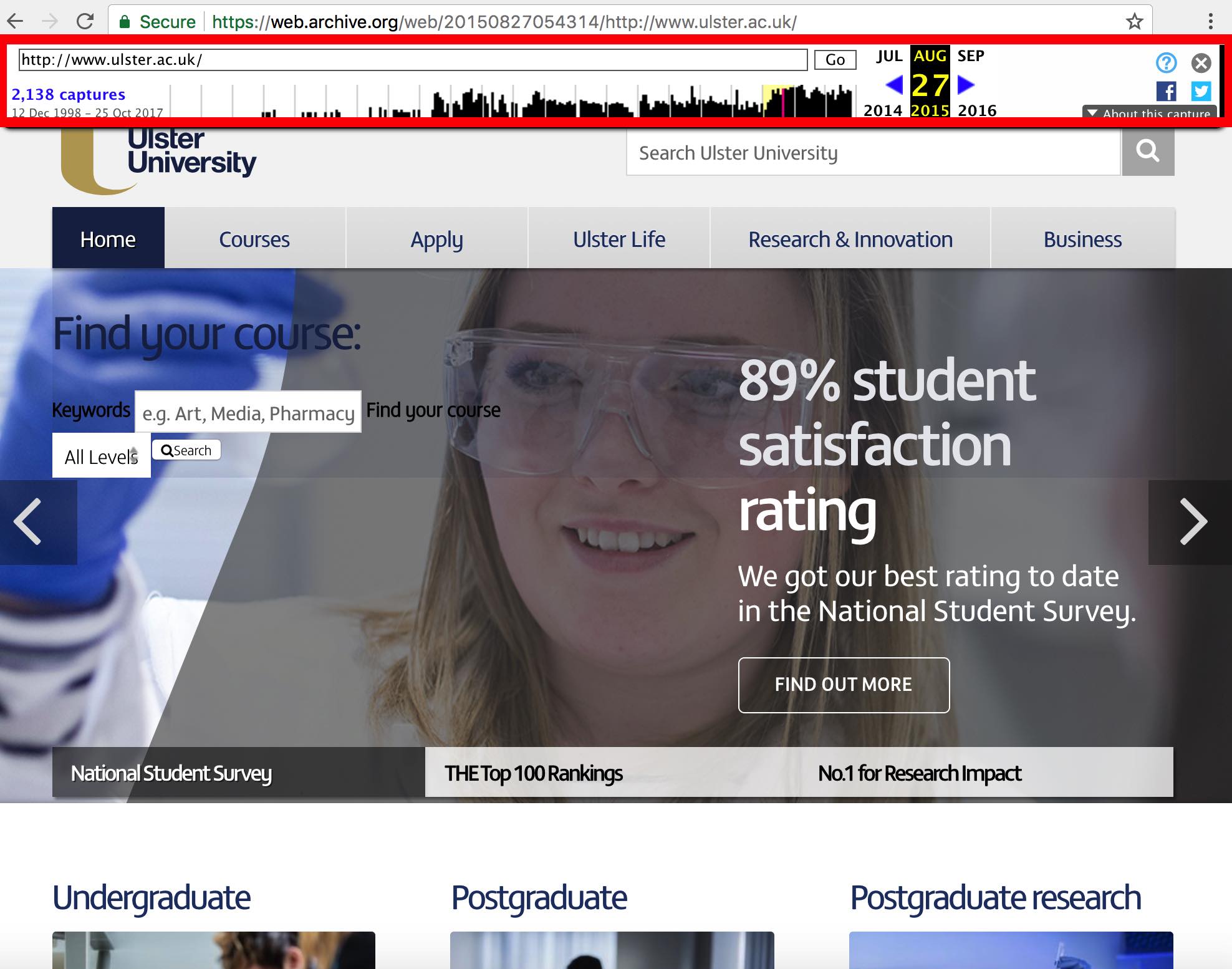}
	}
	\label{fig:banner_ia}
	}	
	\\
	\subfigure[From Proni Archive accessed in 2016]{
	\setlength{\fboxsep}{0pt}%
	\fbox{
	\includegraphics[width=2.7in, height=2.2in ]{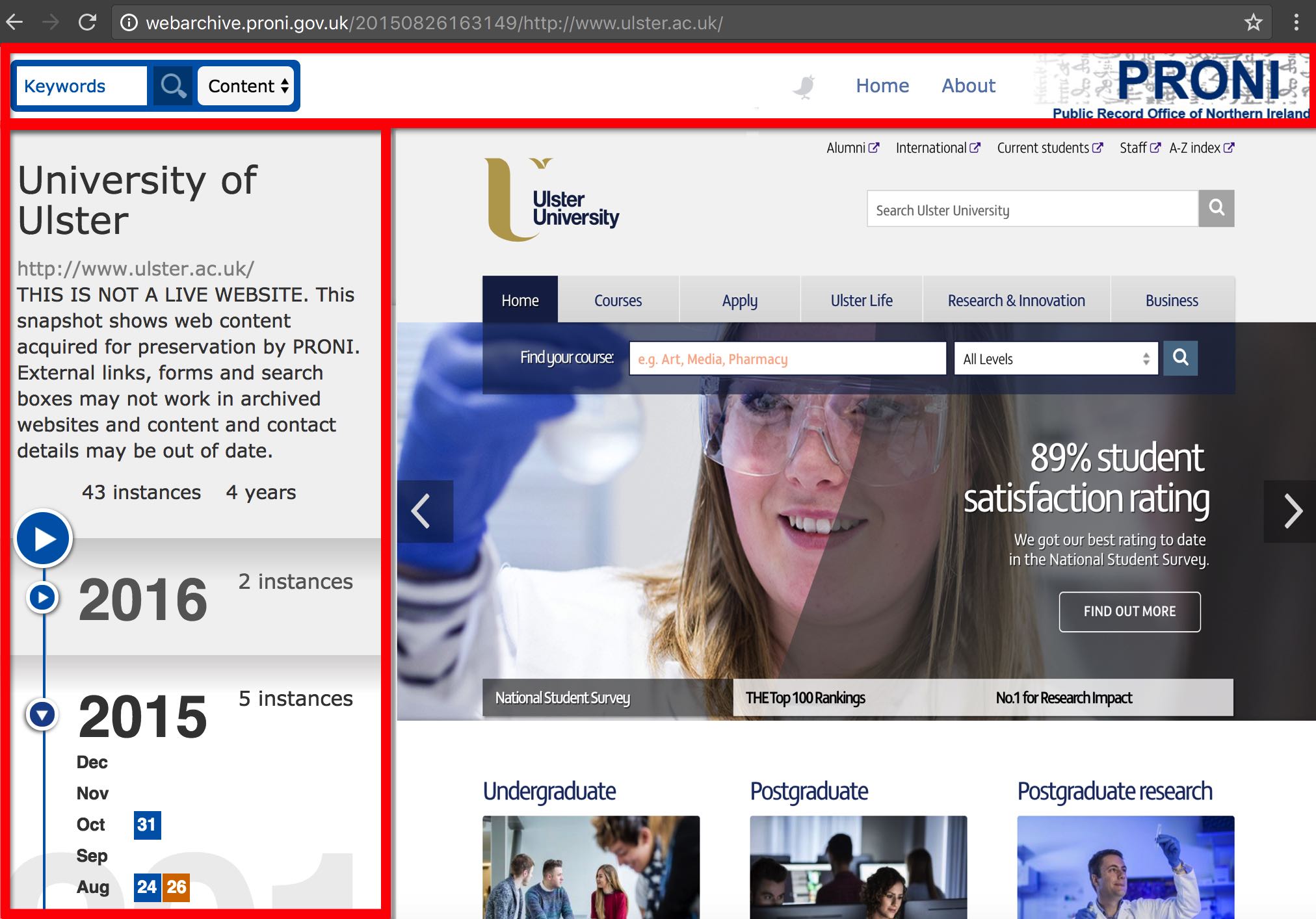}
	}
	\label{fig:banner_proni}
	}	
	\hspace{0.2cm}
	\subfigure[Same memento accessed in 2017]{
	\setlength{\fboxsep}{0pt}%
	\fbox{
	\includegraphics[width=2.7in, height=2.2in]{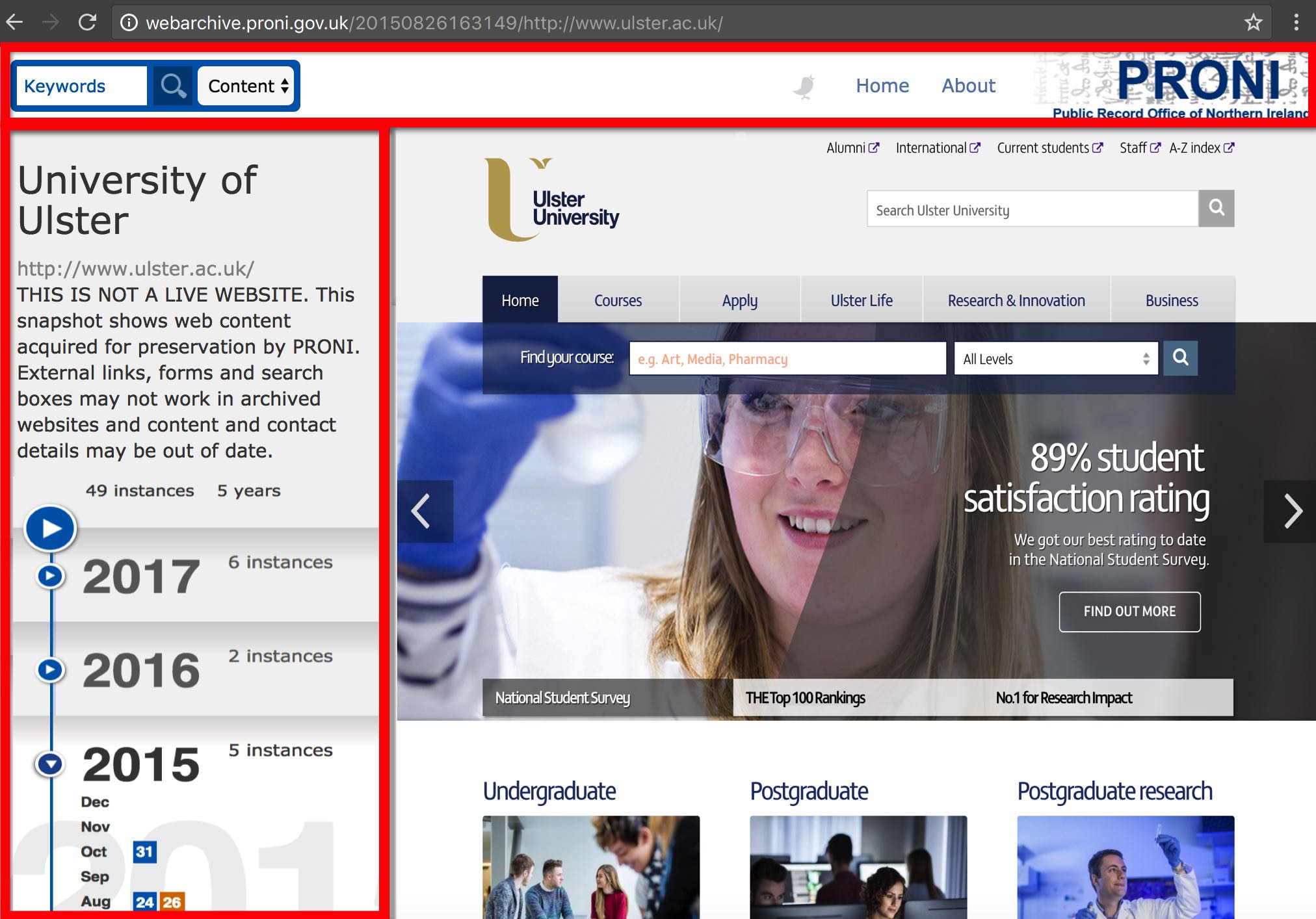}
	}
	\label{fig:banner_proni_new}
	}	
	\hspace{-0.2cm}
\caption{Archive-specific content (marked in red).}
\label{fig:archive-specific-content}
\end{figure*}

Figure \ref{fig:archive-specific-content} shows examples of archive-specific content which are not part of the original content when the memento was initially created. The Internet Archive's banner in  Figure \ref{fig:banner_ia} indicates that there are 2,138 mementos available in the archive for the web page\footnote{\url{http://www.ulster.ac.uk/}}. By hovering the mouse over the banner and clicking on a specific date, the corresponding web page will be displayed in the browser. Figure \ref{fig:banner_proni} presents a different visualization tool to navigate through the archived versions of Ulster University's website\footnote{\url{http://webarchive.proni.gov.uk/20150826163149/http://www.ulster.ac.uk/}}. 

We can identify archive-specific content in archives which support the Memento protocol \cite{memento:rfc}. As shown in Figure \ref{img:donotnego}, archives should respond with the HTTP ``Link'' header containing ``\url{http://mementoweb.org/terms/donotnegot}\\\url{iate}'' and ``\texttt{rel="type"}'' to requests for resources which are not mementos and are excluded from content negotiation based on the time dimension. 

\begin{figure}[!t]
\centering

\begin{Verbatim}[commandchars=\\\{\}]

\userinput{\%} curl -I https://www.webarchive.org.uk/wayback/archive/images/
toolbar/wayback-toolbar-logo.png

HTTP/1.1 200 OK
Date: Mon, 06 Nov 2017 22:35:53 GMT
Server: Apache-Coyote/1.1
\userinput{Link: <http://mementoweb.org/terms/donotnegotiate>; rel="type"}
Accept-Ranges: bytes
ETag: W/"4549-1486118270000"
Last-Modified: Fri, 03 Feb 2017 10:37:50 GMT
Content-Type: image/png
Content-Length: 4549
Content-Language: en

\end{Verbatim}
\vspace{-0.5em}
\caption{One way to identify archive-specific resources is to look at the HTTP Response header ``Link'' that contains ``http://mementoweb.org/terms/donotnegotiate''.}
\label{img:donotnego}
\end{figure}

We want to avoid including archive-specific content in hash calculations for two reasons. First, as mentioned, this type of content does not belong to the content of an original page. Second, resources such as the Wayback Machine's banner in Figure \ref{fig:banner_ia} and the sidebar inserted by Proni's archive in Figure \ref{fig:banner_proni} and Figure \ref{fig:banner_proni_new} are expected to change over time due to updates in archive-specific software (e.g., the Wayback Machine's software). In addition, archive-specific resources may carry dynamically-generated data corresponding to the current state of an archive. For example, the sidebar in Figure \ref{fig:banner_proni} lists all years in which mementos are available. In this case, if a user accesses the memento in 2016, all years which have mementos until 2016 will be part of the information in the sidebar. On the other hand, accessing the same memento in 2017 will result in having a new updated sidebar to include 2017's mementos. Another example of dynamically-generated information is the number of available mementos displayed in the Internet Archive's banner and Proni's sidebar. Thus, we need to avoid including these archive-specific resources when calculating hashes.

\vspace{.5cm}
\begin{center}
\tikzstyle{background rectangle}=[thin,draw=black]  
\begin{tikzpicture}[show background rectangle]
\node[align=justify, text width=10.5cm, inner sep=1em]{
Resources added by archives are not part of the original content and should not be included in the hash calculation.
};
\node[xshift=1ex, yshift=-.7ex, overlay, fill=white, draw=white, above 
right] at (current bounding box.north west) {
\textbf{Requirement 3}: Avoid archive-specific resources
};
\end{tikzpicture} 
\end{center}

\vspace{1em}
\textbf{Extracting ``raw'' content of archived web pages:}
Some archives provide an API or other mechanism to allow users obtain the ``raw'' content of a memento (i.e., the original content without any modification). For example, archives that operate OpenWayback send back raw content when receiving an HTTP request with a URI-M containing ``id\_'' appended to a timestamp. For example
\begin{center}
\begin{tabular}{l}
  \url{https://web.archive.org/web/20100923005105id_/http://www.cnn.co}
  \\
  \url{m:80/}
\end{tabular}
\end{center}

{\raggedleft{}will return the raw content of the memento}

\begin{center}
\begin{tabular}{l}
  \url{https://web.archive.org/web/20100923005105/http://www.cnn.com:8}
  \\
  \url{0/}
\end{tabular}
\end{center}

\begin{figure*}
\centering 
	\subfigure[Requesting the CNN archived page without including id\_ option in the URI\-M: https://web.archive.org/ web/20100923005105/http://www.cnn.com:80/.]{
	\setlength{\fboxsep}{0pt}%
	\fbox{
	\includegraphics[scale=0.15]{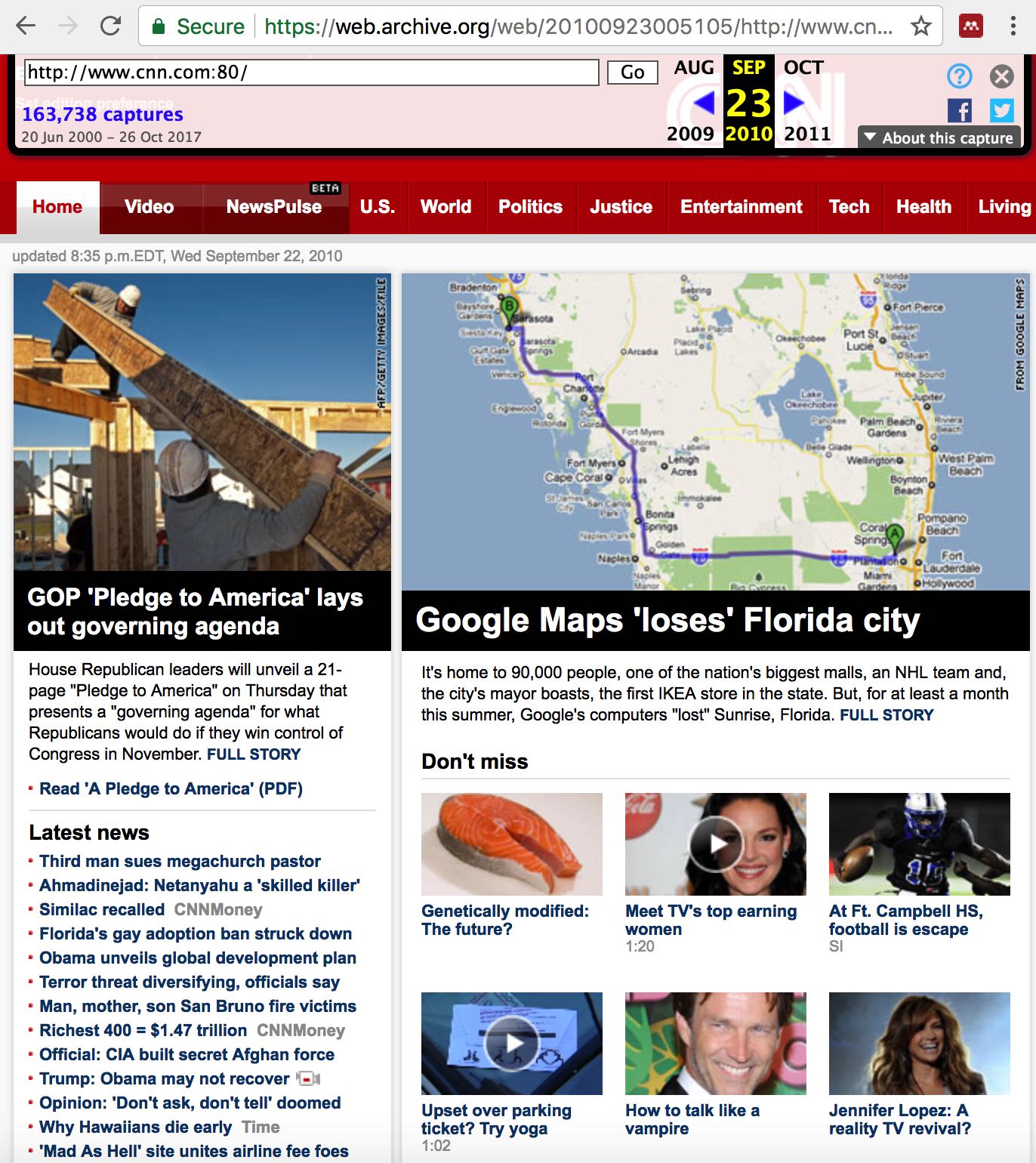}
	}
	\label{img:cnn_rewritten}
	}
	\subfigure[Requesting the raw content of the CNN archived page using id\_ option: https://web.archive.org/web/ 20100923005105id\_/http://www.cnn. com:80/.]{
	\setlength{\fboxsep}{0pt}%
	\fbox{
	\includegraphics[scale=0.15]{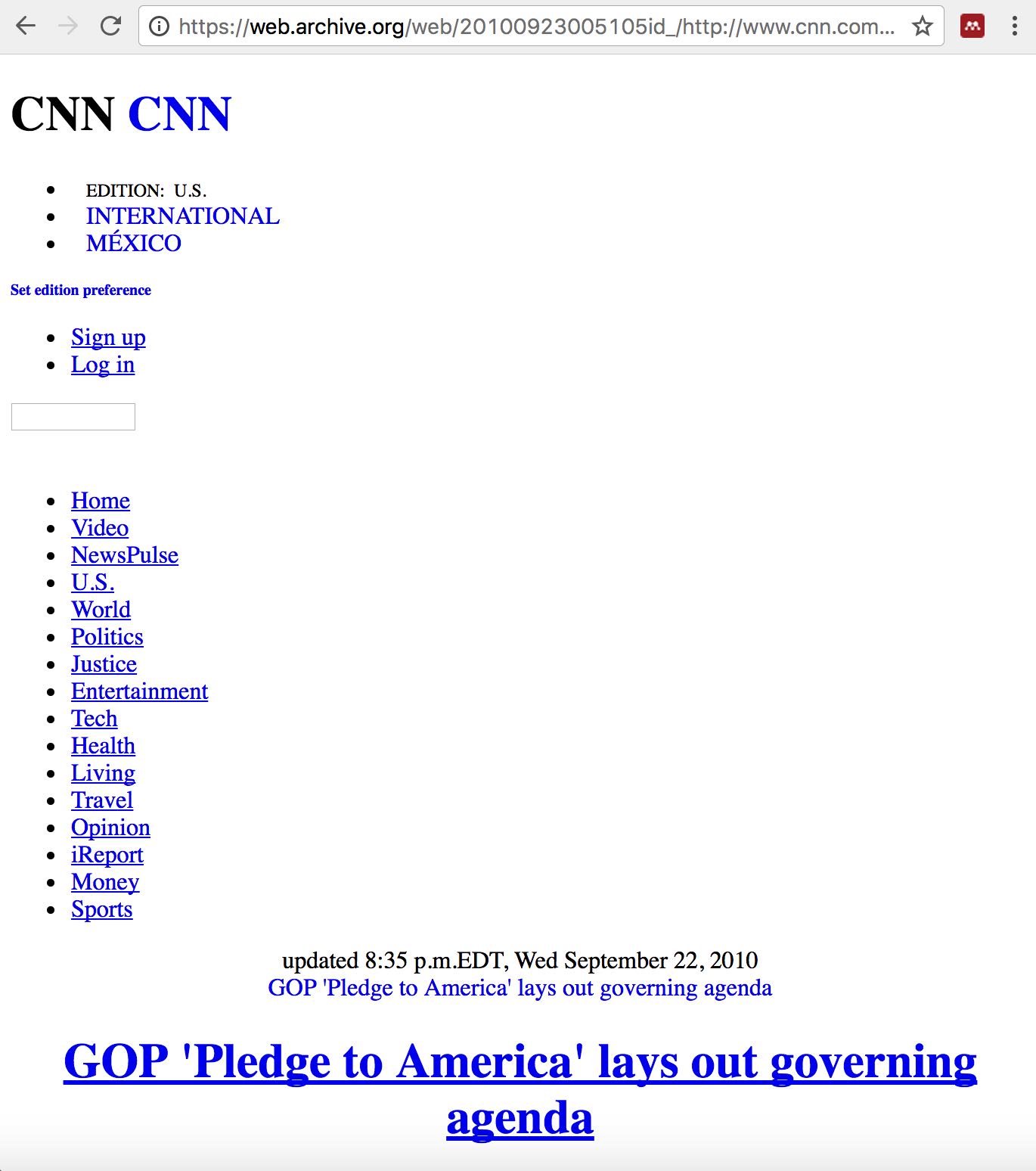}
	}
	\label{img:cnn_raw}
	}	
\caption{The archived web page vs. its raw content.}
\label{fig:cnn_raw_rewritten}
\end{figure*}

Even though using id\_ is beneficial for raw content retrieval, it might cause issues, such as links to resources constructing a memento not being rewritten to point to an archive, which prevents the memento from being replayed appropriately as Figure \ref{fig:cnn_raw_rewritten} shows.

\subsection{Excluding any resources from the live web} \label{live}

Archives rewrite all links of embedded resources of a memento to point to the archive, yet some URIs are not rewritten because they are produced dynamically, for example, by events triggered by client-side JavaScript. Resources specified by such links are often retrieved from the live web \cite{zombie10}.  Web resources from the live web are expected to either change or disappear. Thus, we want to avoid such resources in computing a memento's hash. 

\begin{figure*}[!t]
\centering 
	\subfigure[Accessing the archived web page http://web.archive.org/web/20110901233330/reuters.com page before the ``Archive-Escapes'' attack.]{
	\setlength{\fboxsep}{0pt}%
	\fbox{
	\includegraphics[scale=0.4]{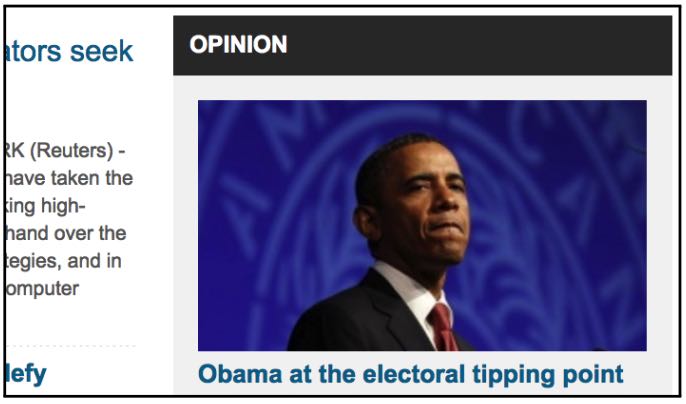}
	}
	\label{img:archive_escapes_before}
	}
	\subfigure[Accessing the same archived page after the ``Archive-Escapes'' attack.]{
	\setlength{\fboxsep}{0pt}%
	\fbox{
	\includegraphics[scale=0.4]{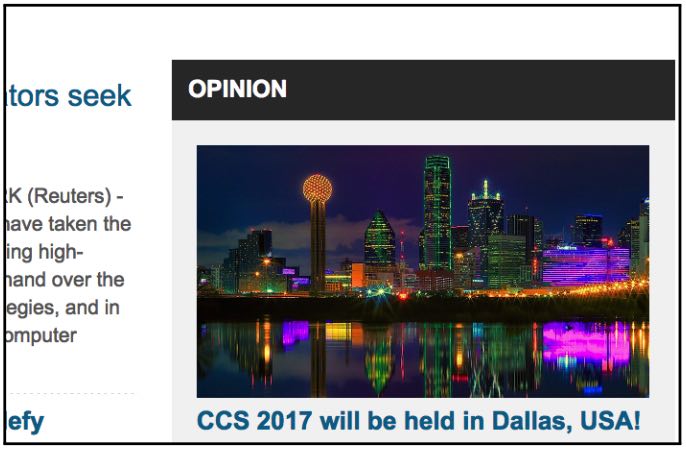}
	}
	\label{img:archive_escapes}
	}	
\caption{A proof of concept of ``Archive-Escapes'' attack -- a request to a live web resource is made when reconstructing a memento (from \cite{lerner2017rewriting}).}
\label{fig:archive_escape_example}
\end{figure*}

Lerner et al. \cite{lerner2017rewriting} explore the web archiving ``Archive-Escapes'' vulnerability which occurs when requesting live web resources as part of constructing a memento. This may lead to changing a user's view when browsing a memento. The authors show a proof of concept implementation (Figure \ref{img:archive_escapes}) of the Archive-Escapes attack. Malicious code is injected in the live web resource

\begin{center}
\url{http://cdn.projecthaile.com/js/trb-1.js}
\end{center}

{\raggedleft{}By requesting this resource as a part of constructing the memento} 

\begin{center}
\url{http://web.archive.org/web/20110901233330/reuters.com}
\end{center}

 {\raggedleft{}it causes a user's view to be completely controlled by the malicious code. This leads to the need for a new requirement.}

\vspace{.5cm}
\begin{center}
\tikzstyle{background rectangle}=[thin,draw=black]  
\begin{tikzpicture}[show background rectangle]
\node[align=justify, text width=10.5cm, inner sep=1em]{
No resource located on the live web should be part of the hashing process.
};
\node[xshift=1ex, yshift=-.7ex, overlay, fill=white, draw=white, above 
right] at (current bounding box.north west) {
\textbf{Requirement 4}: No resources from the live web
};
\end{tikzpicture} 
\end{center}

\subsection{Archived web pages might be served from a cache} \label{cache}

Web archives might use a cache in order to improve performance by speeding up  subsequent requests. The Wayback Machine's HTTP Response header ``X-Page-Cache'' indicates whether delivered content is from the cache (``X-Page-Cache: HIT'') or not (``X-Page-Cache: MISS''). Although caching has powerful benefits, the returned content might not reflect what actually is in the archive. Figure \ref{fig:ia_cache_miss} shows that content is not served from the cache (i.e., ``X-Page-Cache: MISS''). The issue is that cache HITs produce a risk of calculating the same hash even if the archived page has changed. For example, we issued different HTTP requests for the same memento (Figure \ref{fig:ia_cache__hit_miss}). The first response was actually not from the cache with computed MD5 hash ending in ``7afd3'' while the two responses that follow were from the cache. The MD5 hash value calculated on the content of each of the two responses were identical to the first hash value because the content served from the cache is an exact copy of the content returned upon the first request. Now, because the content on the cache is only stored for a ``short'' period of time (depending on how the caching system is configured) before it is discarded (or updated), the fourth response was not a cache HIT. The archived page seems to be changed since we obtain a different MD5 hash value ending in ``b1059''. Thus, we introduce a new requirement when computing a memento's hash.
\vspace{.5cm}
\begin{center}
\tikzstyle{background rectangle}=[thin,draw=black]  
\begin{tikzpicture}[show background rectangle]
\node[align=justify, text width=10.5cm, inner sep=1em]{
We should avoid considering content returned from a cache as this does not reflect the current content of the archive.
};
\node[xshift=1ex, yshift=-.7ex, overlay, fill=white, draw=white, above 
right] at (current bounding box.north west) {
\textbf{Requirement 5}: Avoid content served from the cache
};
\end{tikzpicture} 
\end{center}

\begin{figure}[!t]
\centering
\begin{Verbatim}[commandchars=\\\{\}]

% curl -i http://web.archive.org/web/20130724144801/http://www.c
nn.com/
HTTP/1.1 200 OK
Server: Tengine/2.1.0
Date: Mon, 02 Oct 2017 11:10:15 GMT
Content-Type: text/html; charset=utf-8
Content-Length: 147311
Connection: keep-alive
X-Archive-Orig-set-cookie: CG=US:CA:San+Francisco; path=/
X-Archive-Orig-expires: Wed, 24 Jul 2013 14:48:55 GMT
X-Archive-Orig-vary: Accept-Encoding
X-Archive-Orig-server: nginx
X-Archive-Orig-last-modified: Wed, 24 Jul 2013 14:47:16 GMT
X-Archive-Orig-connection: close
X-Archive-Orig-cache-control: max-age=60, private
X-Archive-Orig-date: Wed, 24 Jul 2013 14:46:36 GMT
X-Archive-Guessed-Charset: utf-8
Memento-Datetime: Wed, 24 Jul 2013 14:48:01 GMT
Link: <http://www.cnn.com/>; rel="original", <http://web.archi
ve.org/web/timemap/link/http://www.cnn.com/>; rel="timemap"; t
ype="application/link-format", <http://web.archive.org/web/htt
p://www.cnn.com/>; rel="timegate", <http://web.archive.org/web
/20000620180259/http://cnn.com:80/>; rel="first memento"; date
time="Tue, 20 Jun 2000 18:02:59 GMT", <http://web.archive.org/
web/20130723125209/http://www.cnn.com/>; rel="prev memento"; d
atetime="Tue, 23 Jul 2013 12:52:09 GMT", <http://web.archive.o
rg/web/20130724144801/http://www.cnn.com/>; rel="memento"; dat
etime="Wed, 24 Jul 2013 14:48:01 GMT", <http://web.archive.org
/web/20130725162936/http://www.cnn.com/>; rel="next memento"; 
datetime="Thu, 25 Jul 2013 16:29:36 GMT", <http://web.archive.
org/web/20000620180259/http://cnn.com:80/>; rel="last memento"
; datetime="Tue, 20 Jun 2000 18:02:59 GMT"
Content-Security-Policy: default-src 'self' 'unsafe-eval' 'uns
afe-inline' data: archive.org web.archive.org analytics.archi
ve.org
X-App-Server: wwwb-app23
X-ts: ----
X-Archive-Playback: 0
X-location: All
\textbf{X-Page-Cache: MISS}

...
\end{Verbatim}
\vspace{-0.5em}
\caption{Memento is not delivered from the cache as the HTTP Response header ``X-Page-Cache: MISS'' indicates.}
\label{fig:ia_cache_miss}
\end{figure}

\begin{figure}[!t]
\centering
\begin{Verbatim}[commandchars=\\\{\}]

% date
Mon Oct  2 01:15:18 EDT 2017
% curl --silent http://web.archive.org/web/20130724144801/htt
p://www.cnn.com/ | md5
\textbf{477b6d923cbb7bf9675a0d2feb37afd3}


% date
Mon Oct  2 01:16:29 EDT 2017
% curl --silent http://web.archive.org/web/20130724144801/htt
p://www.cnn.com/ | md5
\textbf{477b6d923cbb7bf9675a0d2feb37afd3}


% date
Mon Oct  2 01:19:31 EDT 2017
% curl --silent http://web.archive.org/web/20130724144801/htt
p://www.cnn.com/ | md5
\textbf{477b6d923cbb7bf9675a0d2feb37afd3}


% date
Mon Oct  2 02:10:24 EDT 2017
% curl --silent http://web.archive.org/web/20130724144801/htt
p://www.cnn.com/ | md5
\textbf{dda6a9bf091d412cbdc2226ce3eb1059}

\end{Verbatim}
\vspace{-0.5em}
\caption{The first ``cURL'' request was not served  from the cache (i.e., ``X-Page-Cache: MISS'') while the second and third request were cache HITs. After about an hour, the fourth request was a cache MISS and produces a different hash. This example shows that cache HITs produce the same hash even though the memento might have changed.}
\label{fig:ia_cache__hit_miss}
\end{figure}

\subsection{Archived web pages might be in flux (changes in TimeMaps)} \label{timemaps}

Archived resources constructing a memento may have different creation dates (i.e., Memento-Datetime). For example, the main HTML file of the memento

\begin{center}
\begin{tabular}{l}
  \url{https://web.archive.org/web/20170414182743/https://climate.nasa.g} 
  \\
  \url{ov/vital-signs/carbon-dioxide/}
\end{tabular}
\end{center}

{\raggedleft{}was captured on April 14, 2017, while one of the embedded images was captured on April 13, 2017}

\begin{center}
\begin{tabular}{l}
  \url{https://web.archive.org/web/20170413144604im_/https://climate.nas}
  \\
  \url{a.gov/system/time_series_images/582_co2_2002_09.jpg}
\end{tabular}
\end{center}
A TimeMap \cite{memento:rfc} contains a list of all available mementos in the archive for a particular original resource. For example, the TimeMap of the original resource \url{http://www.bbc.com/} contains a list of 27,770 mementos and is available in the Internet Archive at

\begin{center}
  \url{http://web.archive.org/web/timemap/link/http://www.bbc.com/}
\end{center}

{\raggedleft{}From this list, we selected and downloaded the following memento several times}

\begin{center}
  \url{https://web.archive.org/web/20170807231028/http://www.bbc.com/}
\end{center}

{\raggedleft{}We first downloaded this memento on August 14, 2017. We noticed that the TimeMap}

\begin{center}
\begin{tabular}{l}
  \url{http://web.archive.org/web/timemap/link/http://ichef.bbci.co.uk/}
  \\
  \url{wwhp/144/cpsprodpb/730D/production/_97235492_p05brd0w.jpg}
\end{tabular}
\end{center}

{\raggedleft{}of one of the embedded images was empty. The archive had not yet captured this image and responded with ``404 NOT FOUND'' to a request to the rewritten link (URI-M)}

\begin{center}
\begin{tabular}{l}
  \url{https://web.archive.org/web/20170807231028im_/http://ichef.bbci.}
  \\
  \url{co.uk/wwhp/144/cpsprodpb/730D/production/_97235492_p05brd0w.jpg}
\end{tabular}
\end{center}

We downloaded the same memento on August 21, 2017. We found that the image's TimeMap is no longer empty as it consists of one memento

 \begin{center}
 \begin{tabular}{l}
  \url{https://web.archive.org/web/20170807230527im_/http://ichef.bbci.}
  \\
  \url{co.uk/wwhp/144/cpsprodpb/730D/production/_97235492_p05brd0w.jpg}
\end{tabular}
\end{center}

The hash generated on the composite memento on August 14, 2017 ended in ``288d7'' which is different from the hash generated for the same composite memento downloaded on August 21, 2017, ending in ``80845''. 

Brunelle et al. \cite{brunelle2013evaluation} studied the TimeMaps of 4,000 original resources for three months and concluded that the number of mementos in TimeMaps changes and, in some cases, decreases. This will definitely affect how a memento is constructed, and thus will result in different hash value being generated. In addition. Kelly  et al. \cite{kelly_arXiv2017,kelly_jcdl2017} discovered that the number of mementos in the TimeMap of an original web page may vary depending on the tool or the API used to access the archive (e.g., via the Internet Archive's web interface\footnote{\url{https://archive.org/web/}} and the Internet Archive's CDX API\footnote{The Wayback CDX Server API: \url{https://github.com/internetarchive/wayback/tree/master/wayback-cdx-server}}).

\vspace{.5cm}
\begin{center}
\tikzstyle{background rectangle}=[thin,draw=black]  
\begin{tikzpicture}[show background rectangle]
\node[align=justify, text width=10.5cm, inner sep=1em]{
 Changing TimeMaps could affect the computation of hashes. It might be necessary to estimate when a memento becomes stable within the archive to avoid issues of having different hashes.
};
\node[xshift=1ex, yshift=-.7ex, overlay, fill=white, draw=white, above 
right] at (current bounding box.north west) {
\textbf{Requirement 6}: Be aware of the effect of changing TimeMaps
};
\end{tikzpicture} 
\end{center}
	
\subsection{Dynamic content} \label{dynamic}

\begin{figure*}
\centering 
	\subfigure[Accessing https://web.archive.org/web/20130530221910/http://www.cnn.com/ on September 21, 2017 at 10:12 AM (Rainy/thunder icon).]{
	\setlength{\fboxsep}{0pt}%
	\fbox{
	\includegraphics[scale=.33]{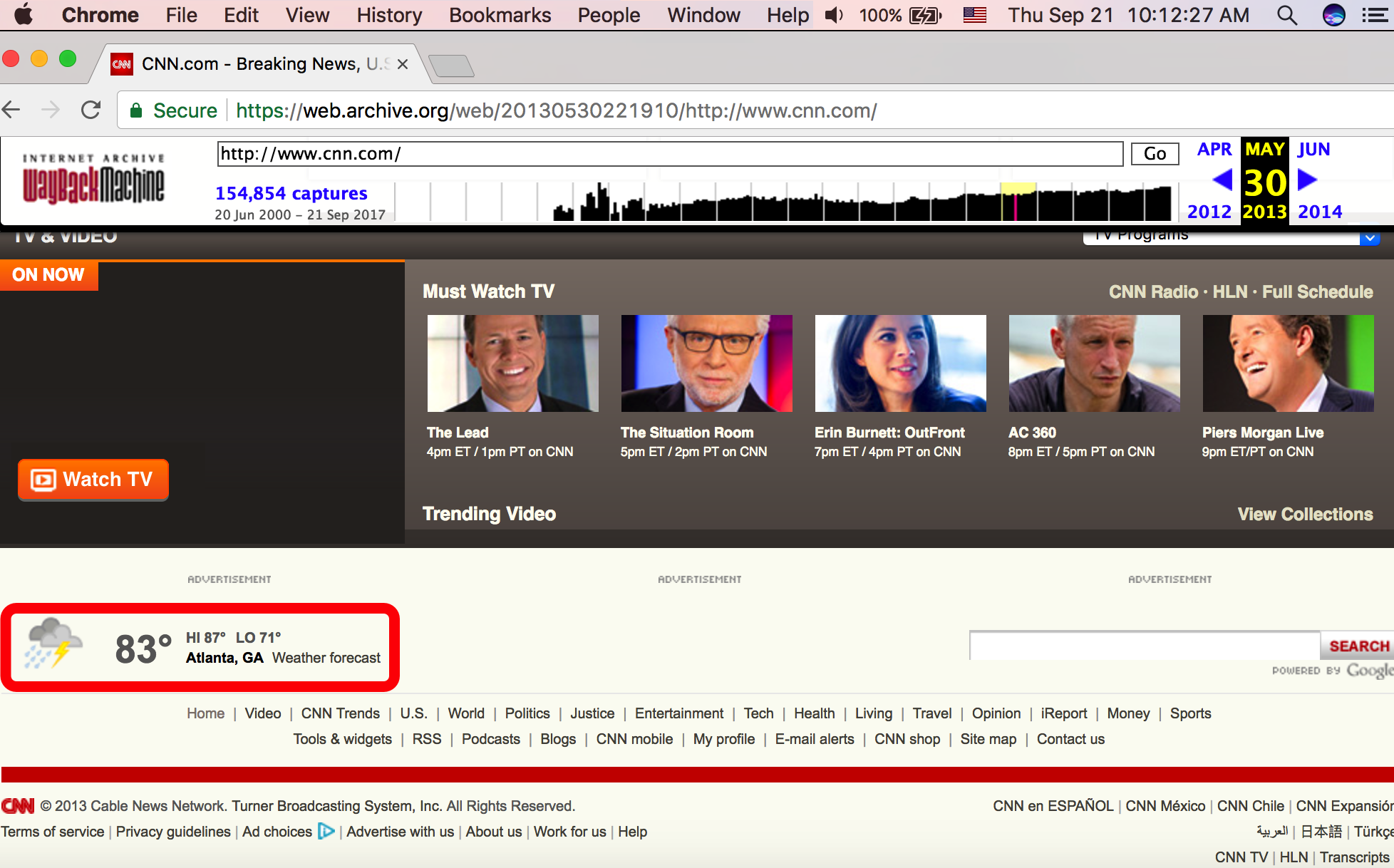}
	}
	\label{img:1_cnn}
	}
	\subfigure[Reloading the same memento at a different time may produce a different icon (Cloudy icon).]{
	\setlength{\fboxsep}{0pt}%
	\fbox{
	\includegraphics[scale=.33]{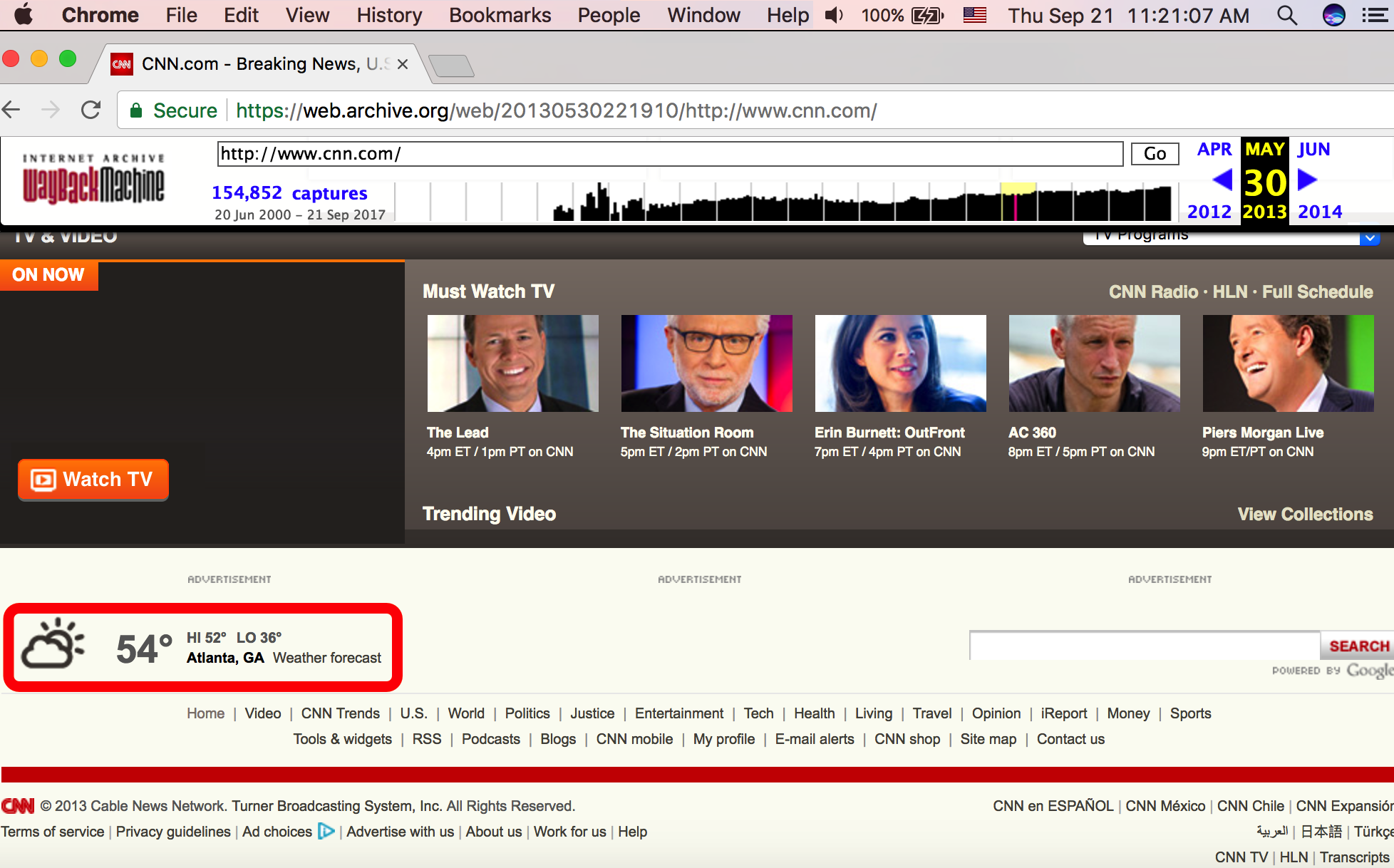}
	}
	\label{img:2_cnn}
	}	
\caption{An example that shows how randomly generated values might affect the hashing process.}
\label{fig:cnn_1_2}
\end{figure*}

Some values are generated on the client-side by JavaScript code which may result in having random or different values each time the JavaScript code is executed. Considering resources with random values may cause inconsistencies in hash calculation. Figure \ref{fig:cnn_1_2} shows a ``rainy/thunder'' icon when the memento:

\begin{center}
 \url{https://web.archive.org/web/20130530221910/http://www.cnn.com/}
 \end{center}
 
 {\raggedleft{} was accessed on September 21, 2017. Reloading the same memento in the browser, we noticed that the icon changed to be ``cloudy''. This happens because the URI to the icon is generated by JavaScript, which involves retrieving the current datetime.}

\vspace{.5cm}
\begin{center}
\tikzstyle{background rectangle}=[thin,draw=black]  
\begin{tikzpicture}[show background rectangle]
\node[align=justify, text width=10.5cm, inner sep=1em]{
Any resources discovered to have randomly generated values should not be a part of the computation of hashes.
};
\node[xshift=1ex, yshift=-.7ex, overlay, fill=white, draw=white, above 
right] at (current bounding box.north west) {
\textbf{Requirement 7}: Avoid using dynamic content in hash calculations 
};
\end{tikzpicture} 
\end{center}

\subsection{Changes in HTTP Response headers} \label{hdrs}

We should include values of important HTTP Response headers in hash computation. For example, we encountered in Figure \ref{fig:ia_cache_miss} a scenario where the value of the HTTP Response header ``X-Page-Cache'' refers to whether content served from the cache or not. Another example would be the value of the HTTP Response header ``Location''. This header is included in the response when the HTTP response status code is ``3xx''. By hashing the value of this HTTP header, we can identify if mementos are served from different URI-Ms and we want keep track of such behavior. Another important header is ``Content-Type''. For instance, we may know that the content of an image has not been tampered with, but it is just the format of the image that changed to PNG, which causes a different hash. Rosenthal et al. \cite{migrationofwebcontent} implemented a proof-of-concept system that demonstrated how an archive can use HTTP content negotiation to transparently migrate resources from one MIME type to another (e.g., image/gif to image/png).  This can be useful if a format type becomes obsolete (as recently happened with Adobe Flash \cite{flash}) or otherwise legally encumbered (as happened with GIF \cite{gif}).


\vspace{.5cm}
\begin{center}
\tikzstyle{background rectangle}=[thin,draw=black]  
\begin{tikzpicture}[show background rectangle]
\node[align=justify, text width=10.5cm, inner sep=1em]{
Important HTTP Response headers should be included in the hash computation.
};
\node[xshift=1ex, yshift=-.7ex, overlay, fill=white, draw=white, above 
right] at (current bounding box.north west) {
\textbf{Requirement 8}: Include HTTP Response headers 
};
\end{tikzpicture} 
\end{center}

\section{Conclusions}
In this paper, we emphasize the importance of timestamping archived web pages, as the number of public and private web archives is increasing, and we do not have the same level of trust in all archives (e.g., Michael's Evil Wayback). We showed that the existing blockchain-based timestamping services do not accept URIs. They accept data by value, such as images and text. Being able to reproduce the same hash of a particular archived web page over time is the key part in the process of generating trusted timestamps. Thus, we discussed several difficulties of generating repeatable hashes of archived web pages and introduced several requirements that should be fulfilled when computing hashes. The proposed requirements include:
\begin{enumerate}
\item A generated hash must be repeatable (Section \ref{issues_section})
\item Generate a hash on a composite memento (Section \ref{composite})
\item Exclude archive-specific resources (Section \ref{archive_specific})
\item Avoid resources from the live web (Section \ref{live})
\item Avoid content served from the cache (Section \ref{cache})
\item Changes in TimeMaps might affect the computation of hashes (Section \ref{timemaps})
\item Avoid including dynamic content or randomly generated content (Section \ref{dynamic})  
\item Include selected HTTP Response headers in hash calculation (Section \ref{hdrs})
\end{enumerate}

In future work, we will explore the above requirements by observing a set of archived web pages from several web archives for a period of time. This will help us identify the type of changes in the content of archived resources that might affect generating repeatable hashes.

\section{ACKNOWLEDGEMENTS}
This work is supported in part by The Andrew W. Mellon Foundation (AMF) grant 11600663.

\bibliographystyle{splncs03}
\bibliography{timestamping}

\begin{thebibliography}{10}
\providecommand{\url}[1]{\texttt{#1}}
\providecommand{\urlprefix}{URL }

\bibitem{openwayback}
{OpenWayback} (October 2005), \url{https://github.com/iipc/openwayback/wiki}

\bibitem{kunze2006warc}
{WARC file format} (ISO 28500:2009) (2009),
  \url{http://www.iso.org/iso/iso_catalogue/catalogue_tc/catalogue_detail.htm?csnumber=44717}

\bibitem{pywb}
{PyWb - Web Archiving Tools for All} (December 2013),
  \url{https://github.com/ikreymer/pywb}

\bibitem{ainsworth2014framework}
Ainsworth, S.G., Nelson, M.L., Van~de Sompel, H.: A framework for evaluation of
  composite memento temporal coherence. Tech. Rep. arXiv:1402.0928 (2014)

\bibitem{gwerntimestamping2017}
Branwen, G.: {Easy Cryptographic Timestamping} (December 2015),
  \url{https://www.gwern.net/Timestamping}

\bibitem{iaurils}
Brewster, K.: {Ta Da! The Wayback Machine has 20 billion more URLs added in the
  last 3 months-- Now 585 billion. go @internetarchive go wayback!} (Sep 2017),
  \url{https://twitter.com/brewster_kahle/status/907576646252642304}

\bibitem{zombie10}
Brunelle, J.F.: {2012-10-10: Zombies in the Archives} (2010),
  \url{http://ws-dl.blogspot.com/2012/10/2012-10-10-zombies-in-archives.html}

\bibitem{brunelle2013evaluation}
Brunelle, J.F., Nelson, M.L.: {An evaluation of caching policies for Memento
  TimeMaps}. In: Proceedings of the 13th ACM/IEEE Joint Conference on Digital
  Libraries (JCDL). pp. 267--276. ACM (2013)

\bibitem{flash}
Collins, K.: {How Adobe Flash, once the face of the web, fell to the brink of
  obscurity and why it's worth saving} (December 2016),
  \url{https://qz.com/863467/how-adobe-flash-once-the-face-of-the-web-fell-to-the-brink-of-obscurity-and-why-its-worth-saving/}

\bibitem{costa2017}
Costa, M., Gomes, D., Silva, M.J.: The evolution of web archiving.
  International Journal on Digital Libraries  18(3),  191--205 (2017)

\bibitem{warcgamesgithub2017}
Cushman, J.: {WARCgames} (May 2017),
  \url{https://github.com/harvard-lil/warcgames}

\bibitem{cushman2017}
Cushman, J., Kreymer, I.: {Thinking like a hacker: Security Considerations for
  High-Fidelity Web Archives} (May 2017),
  \url{http://labs.rhizome.org/presentations/security.html}

\bibitem{eltgrowth2009best}
Eltgrowth, D.R.: {Best evidence and the Wayback Machine: toward a workable
  authentication standard for archived Internet evidence}. Fordham Law Review
  78,  181 (2009)

\bibitem{gif}
{Free Software Foundation}: {Why There Are No GIF Files on GNU Web Pages}
  \url{https://www.gnu.org/philosophy/gif.html}

\bibitem{gippusingpaperforweb}
Gipp, B., Meuschke, N., Breitinger, C.: {Using the Blockchain of
  Cryptocurrencies for Timestamping Digital Cultural Heritage}. In: Proceedings
  of the Workshop on Web Archiving and Digital Libraries (WADL) held in
  conjunction with the 16th ACM/IEEE Joint Conference on Digital Libraries
  (JCDL) (2016),
  \url{https://www.gipp.com/wp-content/papercite-data/pdf/gipp2017a.pdf}

\bibitem{gipp2015decentralized}
Gipp, B., Meuschke, N., Gernandt, A.: {Decentralized trusted timestamping using
  the crypto currency Bitcoin}. Tech. Rep. arXiv:1502.04015 (2015)

\bibitem{jones2016rules}
Jones, S.M., Shankar, H.: Rules of acquisition for mementos and their content.
  Tech. Rep. arXiv:1602.06223 (2016)

\bibitem{jonesraw20162}
Jones, S.M., Sompel, H., Nelson, M.L.: {2016-04-27: Mementos in the Raw}
  (2016),
  \url{http://ws-dl.blogspot.com/2016/04/2016-04-27-mementos-in-raw.html}

\bibitem{kelly_jcdl2017}
Kelly, M., Alkwai, L.M., Alam, S., Nelson, M.L., Weigle, M.C., {Van de Sompel},
  H.: {Impact of URI Canonicalization on Memento Count}. In: Proceedings of the
  17th ACM/IEEE Joint Conference on Digital Libraries (JCDL). pp. 303--304.
  Toronto, Canada (June 2017)

\bibitem{kelly_arXiv2017}
Kelly, M., Alkwai, L.M., Nelson, M.L., Weigle, M.C., {Van de Sompel}, H.:
  {Impact of URI Canonicalization on Memento Count}. Tech. Rep.
  arXiv:1703.03302 (March 2017)

\bibitem{Kim_Nowviskie_Graham_Quon_Alliance_2017}
Kim, K., Nowviskie, B., Graham, W., Quon, B., Alliance, D.S.: {Web Archiving in
  the United States: A 2016 Survey} (Sep 2017), \url{osf.io/r5pqk}

\bibitem{webrecorder}
Kreymer‏, I.: {Webrecorder - a web archiving platform and service for all}
  (2015), \url{https://webrecorder.io}

\bibitem{lerner2017rewriting}
Lerner, A., Kohno, T., Roesner, F.: Rewriting history: Changing the archived
  web from the present. In: Proceedings of the 16th ACM conference on Computer
  and Communications Security (CCS) (2017)

\bibitem{Merkle79}
Merkle, R.: {Secrecy, authentication and public key systems}. Ph.D. thesis,
  Stanford University (1979)

\bibitem{mohr2004introduction}
Mohr, G., Stack, M., Ranitovic, I., Avery, D., Kimpton, M.: {An Introduction to
  Heritrix An open source archival quality web crawler}. In: Proceedings of the
  4th International Web Archiving Workshop (IWAW) (2004)

\bibitem{nakamoto2008bitcoin}
Nakamoto, S.: Bitcoin: A peer-to-peer electronic cash system (2008),
  \url{https://bitcoin.org/bitcoin.pdf}

\bibitem{migrationofwebcontent}
Rosenthal, D.S.H., Lipkis, T., Robertson, T.S., Morabito, S.: {Transparent
  Format Migration of Preserved Web Content}. D-Lib Magazine  11(1) (2005)

\bibitem{elizabethshockman2016}
Shockman, E.: {How do we save the Internet for history? This group is trying}
  (2016),
  \url{https://www.pri.org/stories/2016-01-02/why-group-working-hard-make-sure-internet-isnt-lost-history}

\bibitem{sigurdhsson2010incremental}
Sigur{\dh}sson, K.: {Incremental crawling with Heritrix}. In: Proceedings of
  the 5th International Web Archiving Workshop (IWAW) (2005)

\bibitem{jonesraw2016}
Van~de Sompel, H., Nelson, M.L., Balakireva, L., Klein, M., Jones, S.M.,
  Shankar, H.: {2016-08-15: Mementos In the Raw, Take Two} (2016),
  \url{http://ws-dl.blogspot.com/2016/08/2016-08-15-mementos-in-raw-take-two.html}

\bibitem{tofel2007wayback}
Tofel, B.: Wayback for accessing web archives. In: Proceedings of the 7th
  International Web Archiving Workshop (IWAW). pp. 27--37 (2007)

\bibitem{memento:rfc}
{Van de Sompel}, H., Nelson, M.L., Sanderson, R.: {HTTP framework for
  time-based access to resource states -- Memento, Internet RFC 7089} (2013),
  \url{http://tools.ietf.org/html/rfc7089}

\bibitem{nelson:memento:tr}
{Van de Sompel}, H., Nelson, M.L., Sanderson, R., Balakireva, L.L., Ainsworth,
  S., Shankar, H.: {Memento: Time Travel for the Web}. Tech. Rep.
  arXiv:0911.1112 (2009)

\bibitem{wood2014ethereum}
Wood, G.: Ethereum: A secure decentralised generalised transaction ledger.
  Ethereum Project Yellow Paper  151 (2014), \url{https://ethereum.org/}

\bibitem{wright2015decentralized}
Wright, A., De~Filippi, P.: Decentralized blockchain technology and the rise of
  lex cryptographia. In: SSRN Electronic Journal (2015)

\bibitem{webpackages2017}
Yasskin, J.: {Use Cases and Requirements for Web Packages} (2017),
  \url{https://tools.ietf.org/id/draft-yasskin-webpackage-use-cases-00.html}

\bibitem{perma}
Zittrain, J., Albert, K., Lessig, L.: Perma: Scoping and addressing the problem
  of link and reference rot in legal citations. Legal Information Management
  14(2),  88--99 (2014)

\end{thebibliography}

\end{document}